\documentclass[a4paper,fleqn,usenatbib]{mnras}

\usepackage{newtxtext,newtxmath}

\usepackage[T1]{fontenc}
\usepackage{ae,aecompl}
\usepackage{pdflscape}

\newcommand{\kms}{km$\;$s$^{-1}$}

\usepackage{graphicx}	
\usepackage{amsmath}	
\usepackage{amssymb}	
\usepackage{multirow}
\usepackage{gensymb}

\title[Multi-state spectroscopy of FO Aqr]{Optical spectra of FO Aquarii during low and high accretion rates}

\author[M. R. Kennedy et al.]{M. R. Kennedy$^{1}$\thanks{Contact e-mail: \href{mailto:kennedy.mark@manchester.ac.uk}{kennedy.mark@manchester.ac.uk}}, P. M. Garnavich$^{2}$, C. Littlefield$^{2}$, T. R. Marsh$^{3}$, P. Callanan$^{4}$, \newauthor
R. P. Breton$^{1}$, T. Augusteijn$^{5}$, R. M. Wagner$^{6}$, R. P. Ashley$^{3}$, M. Neric$^{7}$
\\
$^{1}$Jodrell Bank Centre for Astrophysics, Department of Physics and Astronomy, The University of Manchester, Manchester M13 9P, UK\\
$^{2}$Department of Physics, University of Notre Dame, Notre Dame, IN 46556, USA \\
$^{3}$Department of Physics, University of Warwick, Gibbet Hill Road, Coventry CV4 7AL, UK\\
$^{4}$Department of Physics, University College Cork, Cork, Ireland \\
$^{5}$Nordic Optical Telescope, Apartado 474, E-38700 Santa Cruz de La Palma, Spain \\
$^{6}$Department of Astronomy, Ohio State University, Columbus, OH 43210, USA\\
$^{7}$School of Earth and Space Exploration, Arizona State University, Tempe, AZ 85287-1404, USA
}

\date{Accepted XXX. Received YYY; in original form ZZZ}

\pubyear{2019}

\begin{document}
\label{firstpage}
\pagerange{\pageref{firstpage}--\pageref{lastpage}}
\maketitle

\begin{abstract}
Between May 2016 and September 2018, the intermediate polar (IP) FO Aquarii exhibited two distinct low states and one failed low state. We present optical spectroscopy of FO Aquarii throughout this period, making this the first detailed study of an accretion disc during a low state in any IP. Analysis of these data confirm that the low states are the result of a drop in the mass transfer rate between the secondary star and the magnetic white dwarf primary, and are characterised by a decrease in the system's brightness coupled with a change of the system's accretion structures from an accretion disc-fed geometry to a combination of disc-fed and ballistic stream-fed accretion, and that effects from accretion onto both magnetic poles become detectable. The failed low state only displays a decrease in brightness, with the accretion geometry remaining primarily disc-fed. We also find that the WD appears to be exclusively accretion disc-fed during the high state. There is evidence for an outflow close to the impact region between the ballistic stream and the disc which is detectable in all of the states. Finally, there is marginal evidence for narrow high velocity features in the H$\alpha$ emission line during the low states which may arise due to an outflow from the WD. These features may be evidence of a collimated jet, a long predicted yet elusive feature of cataclysmic variables.
\end{abstract}




\section{Introduction}

\subsection{Cataclysmic variables}
Intermediate polars (IPs), which are a class of cataclysmic variable (CV) that contain a weakly magnetic white dwarf (WD) primary and a low mass companion, are key objects for probing accretion flows in magnetic fields due to accretion geometries which occur in these systems. There are currently thought to be three distinct accretion modes in IPs - a ``disc-fed'' mode, a ``stream-fed'' mode, and a ``disc-overflow'' mode. In the ``disc-fed'' mode, an accretion disc forms around the WD primary but is truncated where the magnetic pressure overcomes the ram pressure in the disc. Material then rises out of the orbital plane and follows ``accretion curtains'' to the closest magnetic pole \citep{Rosen1988}. ``Stream-fed'' systems lack a viscous accretion disc, and the accretion stream from the secondary directly impacts the WD's magnetosphere. A fraction of the stream couples to the WD's field lines and accretes, while the remainder can form a non-Keplerian ring of diamagnetic blobs around the WD, providing an additional source of accreting material (\citealt{Hameury1986}; \citealt{1993MNRAS.261..144K}; \citealt{1999MNRAS.310..203K}; \citealt{Hellier2002}). The third mode, called ``disc-overflow'', involves both of these processes, where the WD is fed by material from the disc and the stream at the same time (\citealt{Lubow1989}; \citealt{Armitage1996}). These three accretion modes give rise to various signatures in the optical and X-ray light curves of IPs (\citealt{Wynn1992}; \citealt{Ferrario1999}). The dominant power in the power spectrum of an IP can be at the spin frequency of the WD ($\omega$), the beat frequency of the system ($\omega-\Omega$; where $\Omega$ is the orbital frequency), or a combination of both  of these signals and their harmonics, depending on the accretion mode.

Complications in the study of these systems arise from the existence of low states. Several IPs have been observed to exhibit fading by up to 2 magnitudes at optical wavelengths. These low states are thought to be related to a decrease in mass transfer from the secondary (the low mass companion) to the primary (the WD), potentially caused by the transit of a starspot across the L1 point on the secondary's surface \citep{Livio1994}. Such low states have been seen in the optical light curves of many IPs, but until recently, none have been studied in real time as the only low states have been identified after the events had ended \citep{Garnavich1988}. 

The lack of any studies of an IP during a low state is unfortunate, as spectroscopic observations of an IP during a low state could reveal parameters about the system which are difficult to determine when a high rate of accretion is occurring. For example, the secondary star may become detectable in the optical spectrum if the accretion structures fade sufficiently, allowing for the secondary stars radial velocity curve to be measured, which in turn may lead to limits on the primary WD mass. Additionally, if accretion in an IP ceases completely during a low state, then Zeeman splitting of lines in the optical spectrum due to the WDs magnetic field may become measurable, allowing for a direct measurement of the WDs magnetic field strength \citep{2005ASSL..332..211W}. 

Finally, spectroscopic observations during a low state can answer questions about how the accretion structures respond to a rapid decrease in the mass transfer rate which, theoretically, should have a detectable effect on the accretion geometry, as the ram pressure in the accretion disc and ballistic stream plays a large role in dictating where the infalling material attaches to the WDs magnetic field.

\subsection{FO Aquarii}
FO Aquarii (hereafter FO Aqr) has had a long history of observations since its discovery in 1979 \citep{Marshall1979}. It is considered the archetype of IPs due to its large amplitude and coherent optical and X-ray pulsations at 20.9 mins. The cause of the 20.9 min modulation in the optical light curve has been attributed to the changing viewing angles of a rotating accretion curtain. This same type of modulation has also been detected in the X-ray light curve of FO Aqr \citep{Evans2004}, confirming this period as the spin period of the WD. The spin period has varied over the course of the last 40 years - initially this period was found to be increasing (\citealt{Shafter1982}; \citealt{1987MNRAS.228..193S}), but by the end of the 1980s it was found to have stabilised \citep{1989ApJ...339..434S}. Then in the mid 1990s, the value of $\dot{P}$ flipped, and the period was found to be decreasing (\citealt{Kruszewski1993}; \citealt{1998PASP..110..415P}; \citealt{2003PASP..115..618W}). In the last few years, measurements have suggested that the system was approaching another critical point, and that the sign of $\dot{P}$ was again about to change \citep{Kennedy2016}. New observations in combination with archival photometry of the system have now confirmed that the WD has been spinning down since late 2014 \citep{littlefield19}.

The orbital period was first recorded by \cite{Patterson1983} to be 4.85 h, while the most precise measurement of the orbital period to date is 0.20205976 d \citep{Marsh1996}. A spectroscopic eclipse (that is, a decrease in the flux observed in emission lines) was first proposed by \cite{Hellier1989}. This spectroscopic eclipse was used as evidence that a grazing eclipse of the outer edges of the accretion disc occurs in FO Aqr. However, this proposed spectroscopic eclipse of various emission lines was subsequently subjected to much debate (\citealt{Martell1991}; \citealt{1992ApJ...391..295M}), before being ruled out by \cite{Marsh1996}. The light curve of the source does show a distinct minimum, which has been called the photometric eclipse in literature. The existence of this photometric eclipse, which was firmly established by \cite{Kennedy2016}, has led to a widely adopted inclination of 65\degree\ for the system. To date, there has been no measurement of the mass or spectral type of the secondary star in the system. The mass ratio is assumed to be close to 0.4 based on observations of the secondary stars in all known CVs \citep{2011ApJS..194...28K}.

\begin{figure}
    \centering
    \includegraphics[width=\columnwidth]{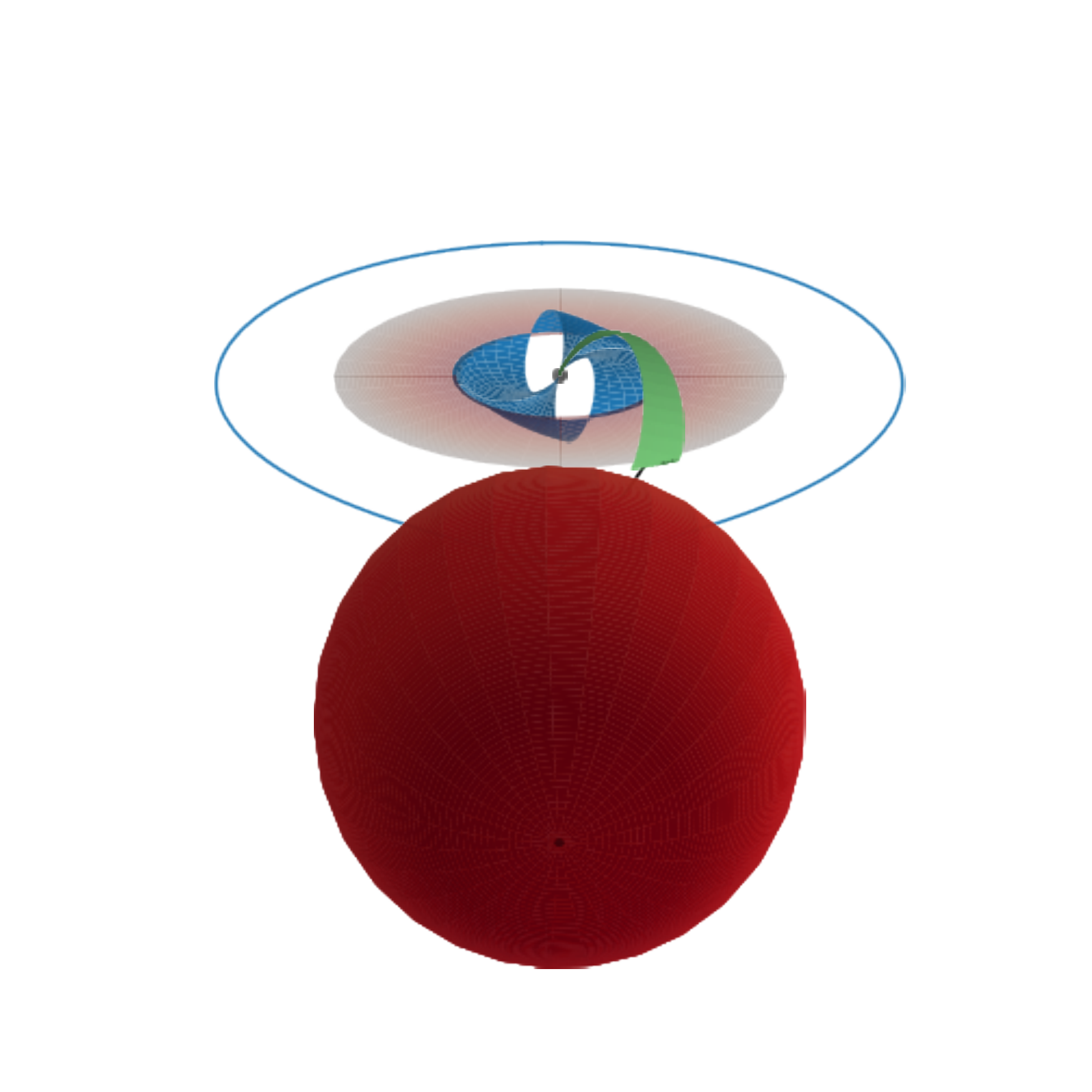}
    \caption{A schematic of what FO Aqr is thought to look like when in its regular high state during the grazing eclipse. Material leaves the secondary (solid, dark red), follows the ballistic stream (black) to the accretion disc. The green lines show material which ``overflows'' the disc and couples directly to the magnetic field of the WD. The inner part of the disc is truncated by the WDs magnetic field, and material begins to follow accretion curtains (shown in blue) onto the surface of the WD (grey). Here, we've adopted a mass ratio of $q=0.4$ and an inclination of 65\degree. The blue circle around the WD denotes the WDs Roche lobe. Note that the sizes of the components are not to scale.}
    \label{fig:schematic_normal}
\end{figure}

In the decade following its discovery, there was significant debate surrounding whether FO Aqr was a pure ``disc-fed'' accretor \citep{Hellier1990}, a pure ``stream-fed'' accretor \citep{Norton1992}, or a ``disc-overflow'' system \citep{Hellier1993}. Recently, it was found that the system shows periods of both pure ``disc-fed'' accretion and periods of ``disc-overflow'' accretion, based on the strength of both the spin and beat frequencies in the power spectrum of data taken using the \textit{Kepler} space telescope \citep{Kennedy2016}. Figure~\ref{fig:schematic_normal} shows a schematic of the system expected at orbital phase 0 (during the eclipse) assuming an inclination of 65\degree.

In the time periods 1923-1953, and 1979-2016, the system was observed to exclusively occupy a bright high state (pre-discovery observations were from the Harvard Plate Collection; \citealt{Garnavich1988}). Recently, it has been discovered that FO Aqr went through several low states between 1953 and 1979 through the study of the APPLAUSE photographic plates \citep{littlefield19}. These low states have gone unnoticed until now.

In 2016, the system was found to be in another low state \citep{littlefield2016c}. The importance of this low state is that it was identified as it was occuring, allowing for detailed observations of the system. Extensive optical photometry and X-ray observations of this low state and its subsequent recovery were carried out which suggested that the low state was caused by a severe drop in the accretion rate, which led to the depletion of the accretion disc in this system (\citealt{littlefield2016c}; \citealt{Kennedy2017}). The most interesting aspect of the low state was the change in which signal was most prominent in the power spectra of the optical and X-ray light curves. The light curve was no longer dominated by the spin period of the WD or beat period of the system, but rather by a signal at half the beat period, 11.2 mins, suggesting that during the low state, we were able to see accretion on to both magnetic poles of the WD. \cite{2017A&A...606A...7H} have suggested that the accretion disc disappeared entirely during the deepest part of the low state. \citet{littlefield19} find circumstantial evidence of disc dissipation in optical photometry but conclude that further theoretical work is necessary to predict photometric behaviours that distinguish discless accretion from stream-overflow accretion. Towards the end of 2016, FO Aqr entered into a recovery state, where the behaviour of its optical light curve matched that of its usual high state behaviour, but its X-ray flux was still fainter than usual \citep{Kennedy2017}.

In August 2017, FO Aqr began to rapidly fade again \citep{2017ATel10703....1L}. Unlike the 2016 low state, during which FO Aqr faded to a V band magnitude of 15.6, the 2017 low state only faded to a V-band magnitude of 14.7 and remained there until the end of that year. Detailed analysis of optical photometry is discussed in detail in \citet{littlefield19}. During the first few months of 2018, FO Aqr seemed to have recovered from the 2017 low state and was hovering about 14.0 mag until, in May 2018, it faded by 0.4 mag \citep{2018ATel11844....1L, littlefield19}. This event was thought to be the beginning of another low state. However, the light curve never faded more than this, and time series analysis show that the beat signal was dominant below V$\sim$14, and the spin was dominant whenever the system was brighter than this.  This low state, which lasted for 5 months, is referred to as a failed low state for the duration of this paper due to the lack of a strong signal at the half beat period and the shallow depth of this event compared to the mean magnitude of FO Aqr during its regular high state.

In this paper, we present optical spectroscopy taken of FO Aqr during various states since 2016. Section~\ref{sec:obs} discusses the telescopes and intruments used. Section~\ref{sec:analysis} presents the data, and introduces the various techniques used to interpret each data set. Section~\ref{sec:2016_low} focuses on data taken during the 2016 low state, Section~\ref{sec:2016_recovery} focuses on data taken during the systems recovery in late 2016, and Section~\ref{sec:2017_low} discusses the data taken during the 2017 low state and the 2018 failed low state. Section~\ref{sec:comparison} compares the spectral shape and emission line behaviour between all the states, alongside a discussion regarding the condition of the accretion disc throughout the various states and potential evidence for detection of an outflow during the low state. Finally, Section~\ref{sec:conc} summarises our conclusions

\section{Observations}\label{sec:obs}
\begin{figure*}
	\includegraphics[width=\textwidth]{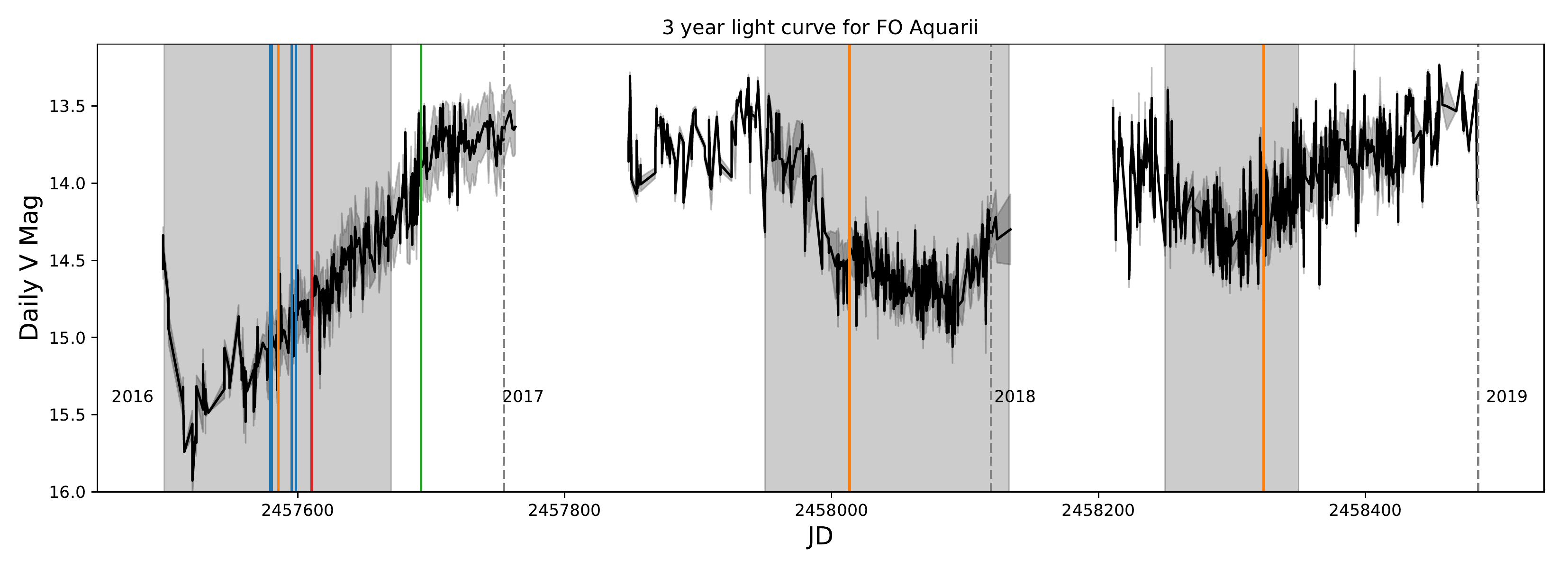}
    \caption{The light curve of FO Aqr between 2016 and the beginning of 2019. Times when spectra were taken using the MDM telescope (blue), NOT (orange), WHT (red), and LBT (green) are marked. The data are a combination of the daily median magnitude of FO Aqr as recorded by observers from the American Association of Variable Star Observers (AAVSO) and from ASAS-SN (\citealt{2014ApJ...788...48S}; \citealt{2017PASP..129j4502K}). The grey outline around the points shows the standard deviation of the magnitudes used to compute the daily magnitude value, while the grey verticle regions mark the 2016 low state, 2017 low state, and 2018 failed low state. The gaps in the data occur when FO Aqr is too close to the Sun to observe.}
    \label{fig:longterm}
\end{figure*}

Figure~\ref{fig:longterm} shows the daily median magnitude of FO Aqr between 2016 and 2019, alongside the nights during which the spectra described below were taken.

\subsection{MDM}
Spectra of FO Aqr were obtained during the 2016 low state using the 2.4~m Hiltner Telescope of the MDM (Michigan-Darthmouth-MIT) Observatory on Kitt Peak. A total of 61 spectra were obtained using the Ohio State Multi--Object Spectrograph (OSMOS) on 2016 July 10, 11, 26, and 29 UT. The blue VPH grism was used with the inner position 1.2\arcsec\ wide entrance slit yielding spectra that cover the range 3980--6860 \AA\ at a dispersion of 0.7 \AA/pixel and a spectral resolution of about $\sim$3.4 \AA. The exposure times were 120~s. The data from July 29 were of particularly poor quality due to heavy clouds throughout the night, leading to very poor S/N in several emission lines. The detector bias was first removed from all the data using the overscan regions of each of the four readout quadrants and then trimmed.  Pixel--to--pixel variations in response (flatfield correction) were removed using the continuous spectra of a quartz--halogen lamp. The combined one dimensional sky--subtracted spectra were then extracted from the two dimensional frames. The spectra were wavelength calibrated using a combination of HgNeArXe lamps obtained in the afternoon before nighttime observations and during the night as well. Relative flux calibration was achieved by exposures of the standard stars BD+33 2642, BD+25 3941, and Feige 110 obtained during the night using the same instrument configuration as the spectra of FO Aqr.

\subsection{Nordic Optical Telescope}
84 spectra of the 2016 low state were also obtained using the Andalucia Faint Object Spectrograph and Camera (ALFOSC) on the 2.5m Nordic Optical Telescope (NOT) on 2016 July 16. ALFOSC was operated in spectro-long-slit mode with grism \# 19, and the CCD was binned 2x2 to reduce the readout time to 9s. The spectra cover the range 4363-6980 \AA\ at a dispersion of 1.2 \AA/pixel with a spectral resolution of about $\sim$5.8 \AA. Each exposure was 40s long. The spectra were wavelength calibrated using \ion{He}/\ion{Ne}{} arc lamp exposures taken directly prior to and after the science data exposures. Flux calibration was carried out using exposures of the standard star Feige 110.

We also obtained spectra of FO Aqr during the 2017 low state on the night of 2017 Sept 18, starting at 22:16:13 (UTC) and lasting for 4 hours. These data were taken when FO Aqr was entering another low state (as detailed in \citealt{2017ATel10703....1L}). For these observations, the spectra were again taken using the ALFOSC instrument but using grism \# 17, which only covers the range 6330-6870 \AA\ with a dispersion of 0.26 \AA/pixel. Each exposure was again 40s long, and the data were wavelength calibrated using \ion{Ne}/\ion{Th}\ion{Ar} arc lamp exposures taken prior to and after the science exposures. The wavelength calibration was found to be accurate to within 1\AA\ for all data by checking the wavelengths of sky lines through out the science exposures.

An additional set of spectra were obtained on the night of 2018 July 24, starting at 23:35:38 (UTC) and lasting for 4.5 hours. These data were taken when FO Aqr was entering a shallow low state \citep{2018ATel11844....1L}, which we call a ``failed'' low state for reasons which will be explored later in the paper. These observations were taken using the same set up as used for the 2017 spectra, and with an exposure time of 40 s.

\subsection{William Herschel Telescope}
FO Aqr was also observed on the night of 2016-08-10 using the Intermediate dispersion Spectrograph and Imaging System (ISIS) mounted on the 4.2m William Herschel Telescope (WHT). ISIS is a dual arm long split spectrograph, allowing for a wide spectral coverage. For these observations the spectra cover a wavelength range of 3700 \AA\ to 5250 \AA\ in the blue arm with a dispersion of 0.44 \AA/pixel, and from 6170 \AA\ to 6940 in the red arm with a dispersion of 0.23 \AA/pixel. Each exposure was 50s long. Since the blue and red arms are not synchronised, the number of spectra taken by each arm differs. There were a total of 263 spectra obtained using the blue arm, and 271 using the red arm. Unfortunately, these spectra could not be flux calibrated due to the lack of observations of a flux standard on the same night. The results presented in this paper are based on the normalised spectra and (in the case of the timing section) continuum-subtracted spectra.

\subsection{Large Binocular Telescope}
The Large Binocular Telescope (LBT) observed FO Aqr using the Multi Object Double Spectrograph (MODS; \citealt{2010SPIE.7735E..0AP}) on 2016 Oct 31, during the systems 2016 recovery state. The term ``recovery state' arises due to the results from \cite{Kennedy2017}, in which it was found that the X-ray flux from FO Aqr was still a factor of 3 lower in October 2016 than its usual high state value. This low X-ray flux suggests that the system had not fully recovered from the preceding low state by the time of the LBT observations. Both instruments were taking data simultaneously, with MODS1 using the SX (left-hand side) mirror and MODS2 using DX (right-hand side) mirror. The start time of both instruments was not synchronised. MODS1 and MODS2 are near-identical in their response and function. In dual grating mode, they cover the wavelength range from 3200 \AA\ up to 1$\micro$m, with a dichroic separating the light into blue and red channels. Both arms had a dispersion of 0.5 \AA/pixel.

A 0.8\arcsec\ segmented long slit was used. The exposure time of each spectrum was 60s, and a total of 100 spectra were taken with each instrument, giving 200 spectra in total. The readout time of the red channel on the MODS instruments is faster than on the blue channel, meaning the red set of spectra finished before the blue set, leading to a slightly different time coverage in the separate channels. The spectra were taken when there was significant cloud coverage visible across the sky, which can be seen as variations of the median flux of the source between sequential exposures.

\section{Analysis}\label{sec:analysis}
We will first show all of the data and discuss the techniques used to create the various plots within the next section. The sub-sections following this will discuss each data set independently, starting with the 2016 low state (Section~\ref{sec:2016_low}), then the 2016 recovery state (Section~\ref{sec:2016_recovery}), and finally a joint discussion on the 2017 low state and the 2018 failed low state (Section~\ref{sec:2017_low}). Section~\ref{sec:comparison} will compare the changes between these states. In all cases, the optical spectra were reduced using tasks in \textsc{Pyraf} and \textsc{Iraf} \footnote{\textsc{Iraf} is distributed by the National Optical Astronomy Observatory, which is operated by the Association of Universities for Research in Astronomy (AURA) under cooperative agreement with the National Science Foundation}.

\begin{figure*}
	\includegraphics[width=\textwidth]{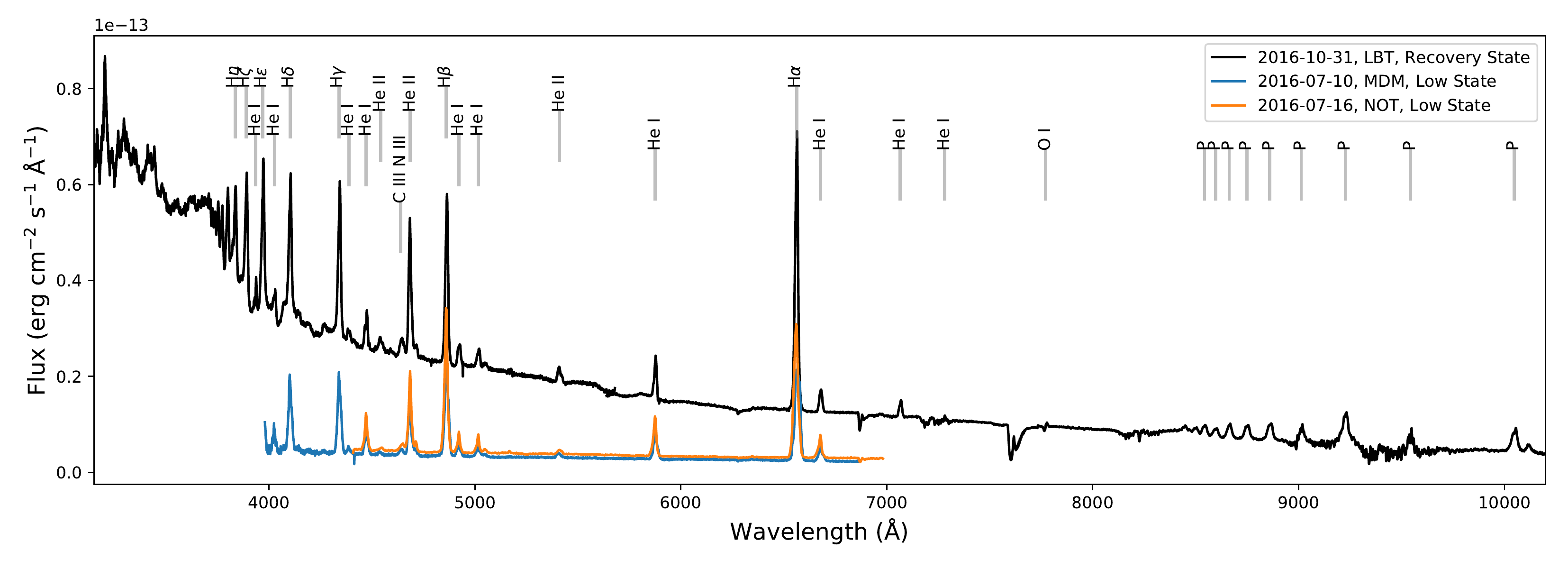}
    \caption{The average spectra taken by the MDM (blue), NOT (orange), and LBT (black) telescopes. Both the MDM and NOT spectra were taken during the 2016 low state, while the LBT spectra were taken in October 2016, during the systems subsequent recovery. The Balmer series of H lines have been marked up to H$\rm{\eta}$, while the visible Paschen series lines have all been marked with a P. The absorption features at 6900, 7200, 7600, 8200, and 9600 \AA\ are telluric features. The weak absorption features are interstellar in origin.}
    \label{fig:average_spectra}
\end{figure*}

The average spectrum of the 2016 low state taken using the MDM telescope on the night of best weather conditions (2016 July 10) is shown in Figure~\ref{fig:average_spectra}, alongside the average of the 2016 low state NOT and 2016 recovery state LBT observations. The LBT spectra were taken 3 months after the MDM and 2016 NOT spectra. At this scale, the two most obvious differences between the average spectrum in the low state (NOT/MDM) and during the recovery state (LBT) is the level and shape of the continuum. Figure~\ref{fig:average_spectra} clearly shows that the continuum level of the average spectrum was significantly lower during the low state than in the recovery state.

\subsubsection*{Orbital Variation}
\begin{figure*}
	\includegraphics[width=0.9\textwidth]{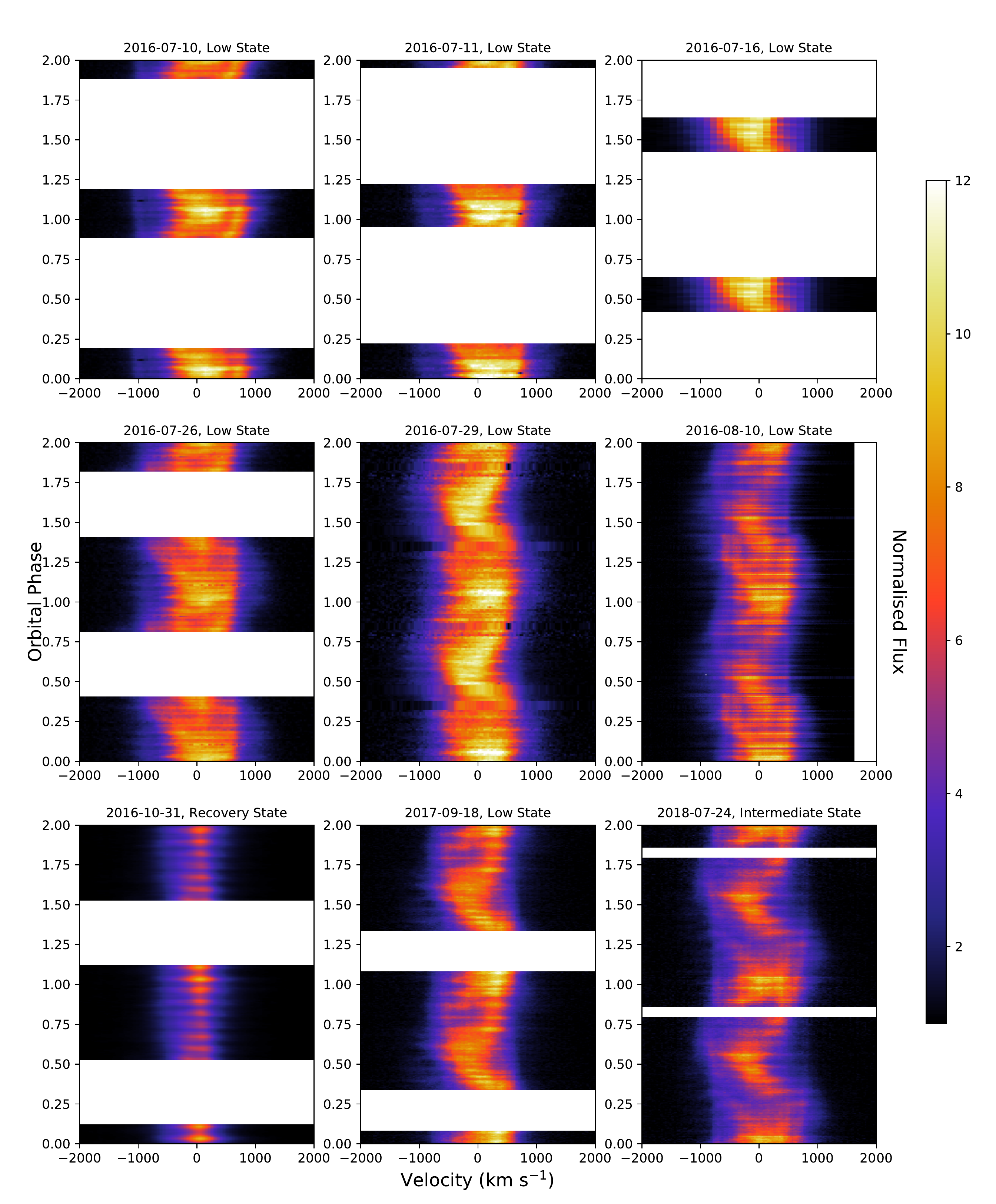}
    \caption{The trailed spectra of H$\alpha$. The data were phased using the ephemeris given in the text. The data from 2018-07-29 is of significantly poorer quality than the other data. The data from 2016-07-10, -11, -16, -26, and -29 were all taken during the 2016 low state, while the data from 2016-10-31 were taken when the system has returned to its normal optical magnitude of $\sim$ 13.6. There are obvious pulsations in each set of data, albeit at different periods (see Figure~\ref{fig:spectrograms} and Figure~\ref{fig:v_r}). The data from 2017-09-18 were taken from the second low state in 2017, and the features in the trailed spectra match those from the 2016 low state remarkably well. The final trailed spectrum from 2018 were taken when FO Aqr was believed to be entering another low state. The data have been repeated over orbital phase for clarity.}
    \label{fig:trailed_ha}
\end{figure*}

\begin{figure*}
	\includegraphics[width=0.9\textwidth]{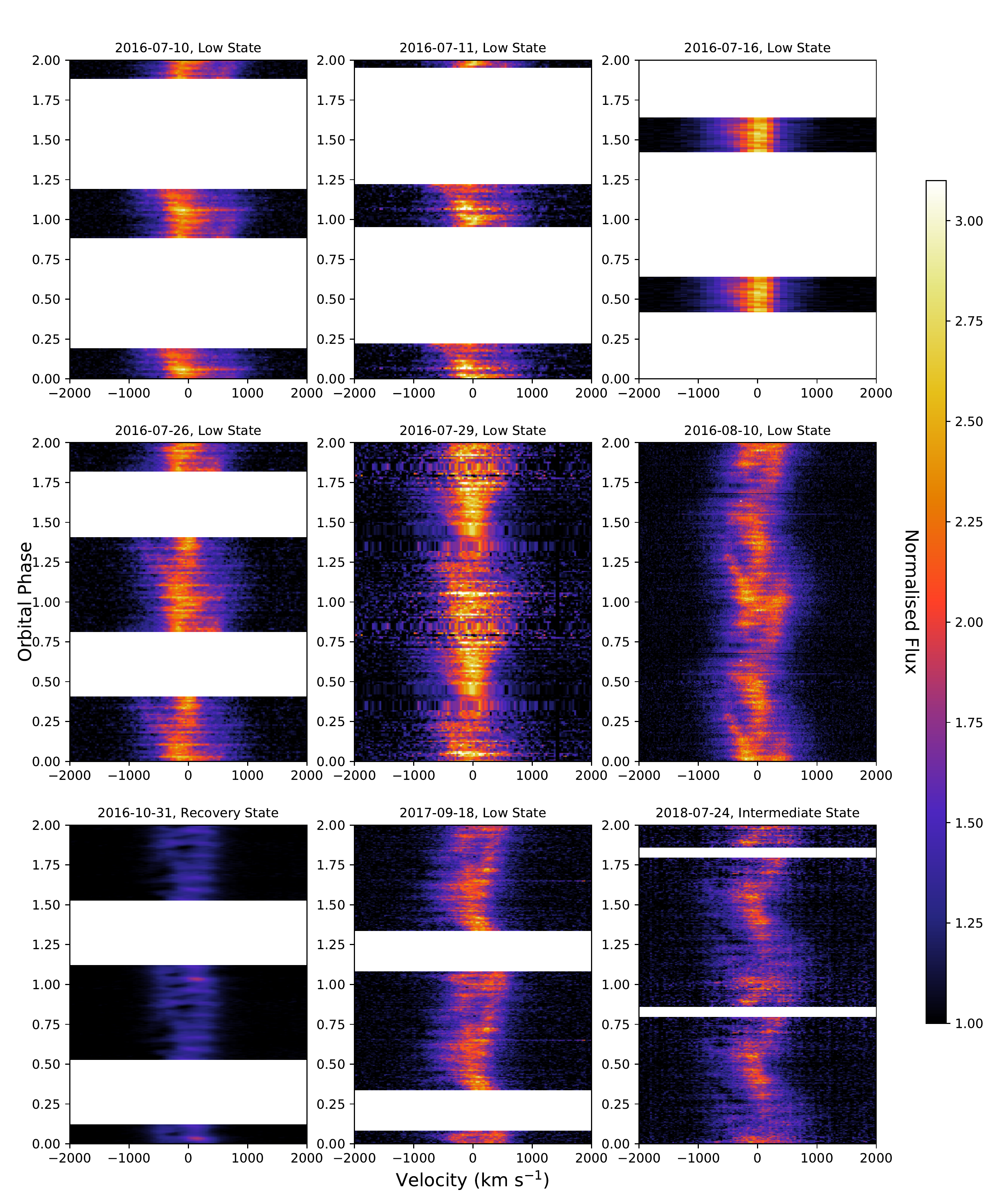}
    \caption{The same as Figure~\ref{fig:trailed_heI}, but for \ion{He}{I} 6678 \AA. The data from 2016-07-29 has been severely affected by clouds, making detection of the line difficult. Data has been repeated over an orbital phase for clarity. Scaling is in normalised flux, which is why the data from 2016-10-31 (when the continuum was significantly stronger) is much fainter.}
    \label{fig:trailed_heI}
\end{figure*}

\begin{figure*}
	\includegraphics[width=0.9\textwidth]{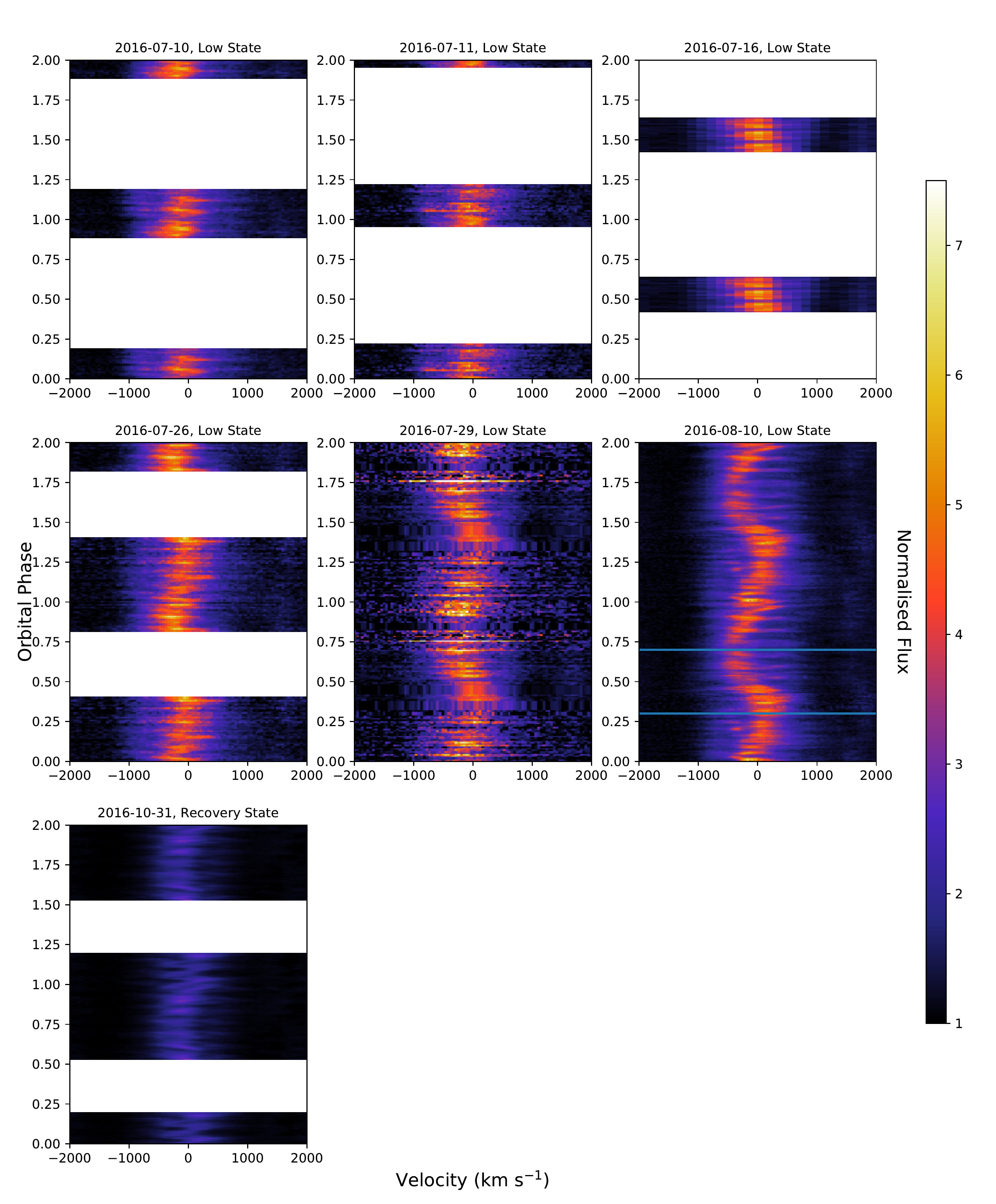}
    \caption{The same as Figure~\ref{fig:trailed_heII}, but for \ion{He}{II} 4686. There are no data from 2017 or 2018 as the grating used in these observations has a very narrow spectral range of 6330-6870 \AA.}
    \label{fig:trailed_heII}
\end{figure*}

To better see the structure of the individual lines, we trailed the spectra taken during each observation (see Figure~\ref{fig:trailed_ha} for H$\alpha$, Figure~\ref{fig:trailed_heI} for \ion{He}{I} 6678\AA, and Figure~\ref{fig:trailed_heII} for \ion{He}{II} 4686\AA). The 2017 NOT and 2018 NOT data are only shown in the H$\alpha$ and \ion{He}{I} panels as the spectral range of these data covered 6330-6870 \AA. All data were phased using the orbital ephemeris of

\begin{equation}
    T(\mathrm{BJD})=2456982.2278(8)+\phi*0.20205976
\end{equation}

where $\phi$ is the orbital phase, and with $\phi=0$ is the time of minimum light in the optical light curve.

\subsubsection*{Spin \& Beat variations} \label{sec:SpinBeat}

\begin{figure*}
	\includegraphics[width=0.9\textwidth]{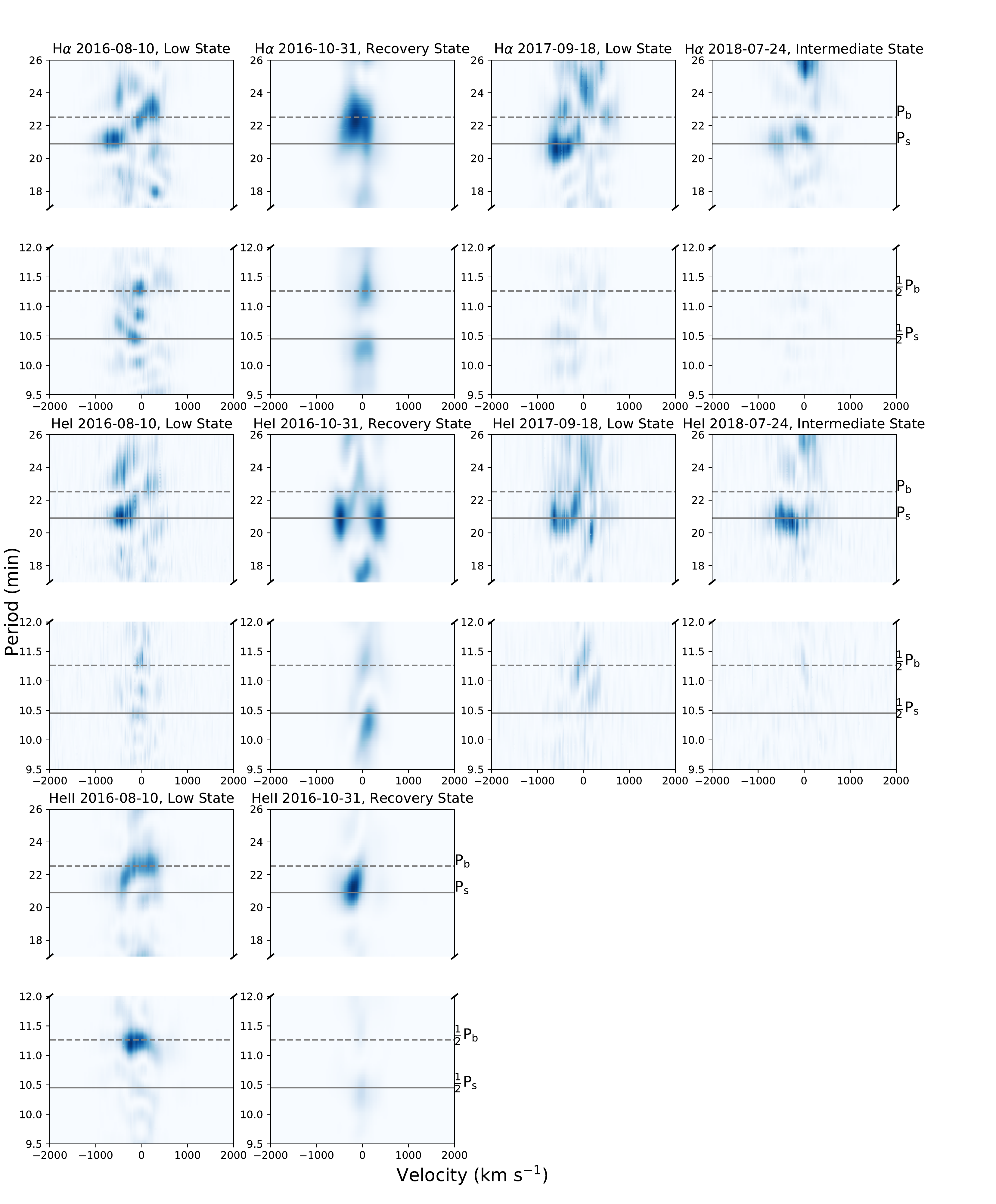}
    \caption{Spectrogram of 3 emission lines during various different states observed in FO Aqr. The far left column shows data from 2016-08-10 (low state), the middle left column shows data from 2016-10-31 (recovery state), the middle right column shows data from 2017-09-18 (low state), and the right column shows data from 2018-07-24 (failed low state). Top row shows data for H$\alpha$, middle row shows data for \ion{He}{I} $\rm{\lambda}$ 6673, and the bottom row shows data for \ion{He}{II} $\rm{\lambda}$ 4686. The spin period and its second harmonic have been marked with solid grey lines, while the beat period and its second harmonic have been marked with a dashed grey line.}
    \label{fig:spectrograms}
\end{figure*}

To study which periods were dominant during each spectroscopic observation of FO Aqr, the Lomb-Scargle periodogram (\citealt{Lomb76}; \citealt{scargle82}) of each wavelength of the spectra (after continuum subtraction) for every observation was taken. These spectrograms (periodograms a function of wavelength) for wavelength ranges around the H$\alpha$, \ion{He}{I}, and \ion{He}{II} lines for several different observations are in Figure~\ref{fig:spectrograms}. 

The combination of these periodicities present in each line makes the creation of trailed spectra folded on any one period very difficult to interpret, as any folded spectra will be contaminated by the other signals present within the line.

\begin{figure*}
	\includegraphics[width=0.9\textwidth]{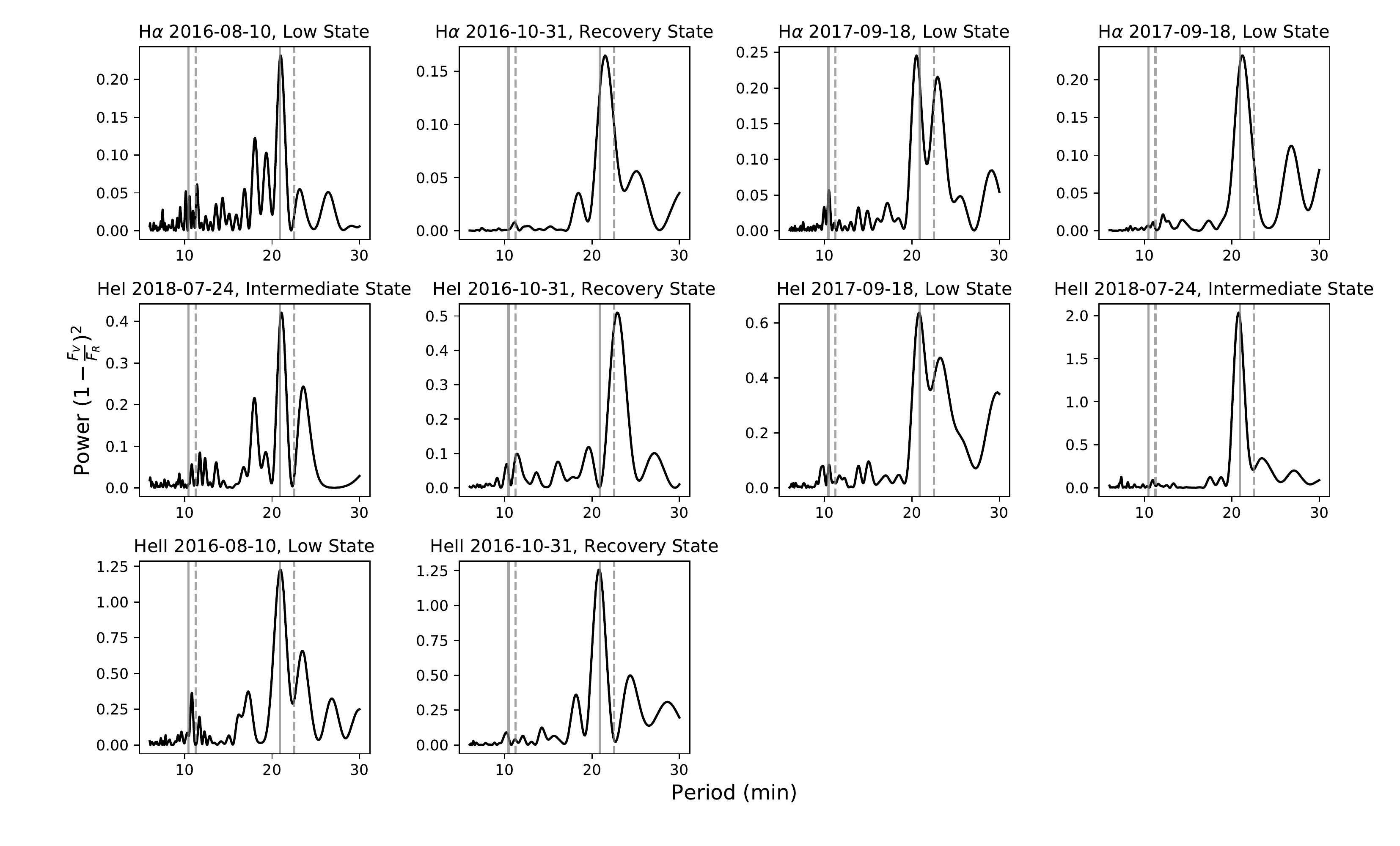}
    \caption{The Lomb-Scargle periodograms of the $V/R$ values for H$\alpha$ (first row), \ion{He}{I} $\lambda$ 6678 \AA\ (second row), and \ion{He}{II} $\lambda$ 4686 \AA\ (third row). The spin period and half-spin period are marked with solid grey lines, while the beat period and half-beat period are marked with a dashed grey line. The highest peaks in these periodograms indicate the dominant periods at which the ratio of the flux at high and low velocities relative to the line centre changes.}
    \label{fig:v_r}
\end{figure*}

While the spectrogram is useful in establishing the time scale for variations within the lines, it is incapable of answering whether the periodicities arise due the flux from all of the line increasing and decreasing over that time scale, or if there are individual components in the lines which have constant flux but move from positive to negative velocities. To answer which of these behaviours is giving rise to the periodicities detected in the periodograms, the ratio of flux at negative velocities compared to the flux at positive velocities (with respect to the line centre) for each spectrum was computed. Hereafter this is referred to as the $V/R$ ratio. A Lomb-Scargle periodogram was then calculated for the $V/R$ values for different accretion states. These periodograms are shown in Figure~\ref{fig:v_r}.

Before proceeding further, it is useful to discuss which accretion geometries are expected to produce distinct signals in the spectrograms and in the periodogram of the $V/R$ ratio. In the following, we assume a single accreting pole is detectable (for two accreting poles, the discussed periods should be halved), and that material in an accretion disc couples to the WDs magnetic field, creating an accretion curtain. We also define orbital phase 0 to be when inferior conjunction of the secondary occurs.

If the emission is coming from the region where the accretion disc is being truncated by the WDs magnetic field, then the resulting accretion curtains will produce a signal at the WD spin period in the spectrogram. This is due to the visible area of the curtain changing as it rotates causing a periodic change in the brightness of the source. It should also produce a signal at the spin period in the $V/R$ ratio as the observed velocity of the material will vary over the spin period, with the emission reaching maximum blue shift when the magnetic pole is pointing away from us, such that material in the curtain is flowing towards us.

Alternatively, the emission could be coming from periodic heating of a fixed structure in the binary (for example the inner face of the secondary star or the bright spot on the edge of the accretion disc) by X-rays generated at the accreting pole of the WD. There should then be a signal in the spectrograms with period equal to how often the accreting pole sweeps across this area, which will be the beat period of the system. If this is the cause of the emission, there should not be any signal in the periodograms of the $V/R$ ratio, as there is no change in material flowing towards or away from the observer over the beat period.

If material is coupling directly between the ballistic stream and the WDs magnetic field, the picture becomes more complicated. We would definitely expect a signal at the beat period of the system in the spectrogram. However it is not clear what signal, if any, should be present in the $V/R$ ratio. Between orbital phases 0.2 and 0.7, the coupling point should be on the far side of the orbit from an observer on Earth, meaning any material which is following field lines from the connection point towards the magnetic pole will be blue shifted, with maximum blue shift reached around phase 0.45 (assuming the connection point leads the secondary star by 0.05 in orbital phase). For orbital phases 0.7 to 1.2, the material following any field lines will be flowing away from us, producing red shifted emission. This should mean the $V/R$ ratio is modulated at the orbital period. However, there may also be power at shorter periods, depending on the radial and azimuthal extent of the region which channels the accretion.

\subsubsection*{Doppler Tomography}
Finally, several of the following data sets were used to create Doppler tomograms. Doppler tomography is a method of constructing approximate images of accretion structures in velocity space by modelling the evolution of a given emission line over a defined period, and was first pioneered by \cite{1988MNRAS.235..269M}. Here, we use the tools developed by \cite{2015A&A...579A..77K}, which rely on code created by \cite{1998astro.ph..6141S}. We create both typical tomograms, where low velocity material lies towards the middle of the tomograms and is not well resolved, and inverted tomograms, where low velocity material lies towards the outer extents of the tomograms and structure is more readily identifiable.

\subsection{2016 Low State} \label{sec:2016_low}

\subsubsection{The optical magnitude of the secondary star}

There are no signs of spectral features from the secondary star in any of the individual spectra, or in the average spectrum. Given the long orbital period of FO Aqr (4.8 hr), the mass ratio is expected to be close to 0.4 \citep{2011ApJS..194...28K}. Assuming that the WD in FO Aqr has a mass of 0.87 M$_{\odot}$ (comparable to the median mass of WDs in CVs; \citealt{2011A&A...536A..42Z}) then the secondary in FO Aqr should have a mass of 0.35M$_{\odot}$. If this star is a main sequence star, then it should be of the M2-M4 spectral type, and have an absolute magnitude between 7-9. Recently, the parallax of FO Aqr was accurately measured by the \textit{Gaia} space telescope \citep{2018A&A...616A...1G}, which places the system at a distance of $514^{+40}_{-15}$ pc from Earth. This distance combined with the expected absolute magnitude of the secondary suggests that the apparent magnitude of the companion should be between 17-18. This is 2 magnitudes fainter than the magnitude of the source during the faintest epoch observed, suggesting that the spectrum was still being dominated by the accretion structures present during the low state.

\subsubsection{Orbital Behaviour}

The trailed spectra show complex behaviour in each of the emission features during the low state. This is particularly true for H$\alpha$. The trailed spectra from 2016-08-10 show at least 3 distinct components which are distinguishable around $\phi \sim0.25$ - a broad component moving with low velocity around the core of the emission line, and 2 narrow components on either side. These components are also visible in the other spectra taken during the 2016 low state, and also in the trailed spectra of \ion{He}{I}, meaning these emission features must be arising in the same place in the binary for both lines.

The wide component, when visible, traces a near S-wave pattern in the trailed spectra. This component is visible in spectra taken during the high state, and has been associated with emission from the secondary star, hot spot, or threading region within the system (\citealt{Hellier1990}; \citealt{Marsh1996}). Additionally, the high state spectra show a strong S-wave absorption feature alongside the main emission component (see Plate 1 of \citealt{Hellier1990} and Figure 8 of \citealt{Marsh1996}). While not as obvious in our data, this component is still present, as highlighted by the blue-to-red moving absorption component visible at phase 0.75 in the trailed spectra of \ion{He}{I} taken on 2016-08-10. This is the same orbital phase at which both \cite{Hellier1990} and \citet{Marsh1996} reported seeing the strongest absorption in the line.

The shape of \ion{He}{II} 4686 \AA\ is simpler, with a single strong S-wave emission component visible in the trailed spectra during the low state, moving from close to 0 km s$^{-1}$ to positive velocities (red shifted), reaching maximum red shift around $\phi \sim$0.3 followed by a maximum blue shift just before $\phi \sim$0.7. The behaviour over the 4.8 hr orbital period is the same as in the trailed spectra shown in \cite{Hellier1990} and \citet{Marsh1996}, with their data showing the same maximum red shift close to orbital phase 0.3 and maximum blue shift close to orbital phase 0.7. However, the emission component which moves rapidly from blue-to-red and red-to-blue over a period of $\sim20$ min seen in Figure 9 of \citet{Marsh1996} is not obvious in the low state data.

\subsubsection{Doppler Tomography}\label{sec:DopTomog}

\subsubsection*{Systemic Velocity}
One of the requirements for creating accurate tomograms is a measurement of a systems systemic velocity ($\gamma$) as an incorrect $\gamma$ can introduce artefacts into tomograms. Previous attempts to measure $\gamma$ for FO Aqr have been restricted by the multiple components visible in trailed plots of H$\beta$. 

\begin{figure*}
	\includegraphics[width=\textwidth]{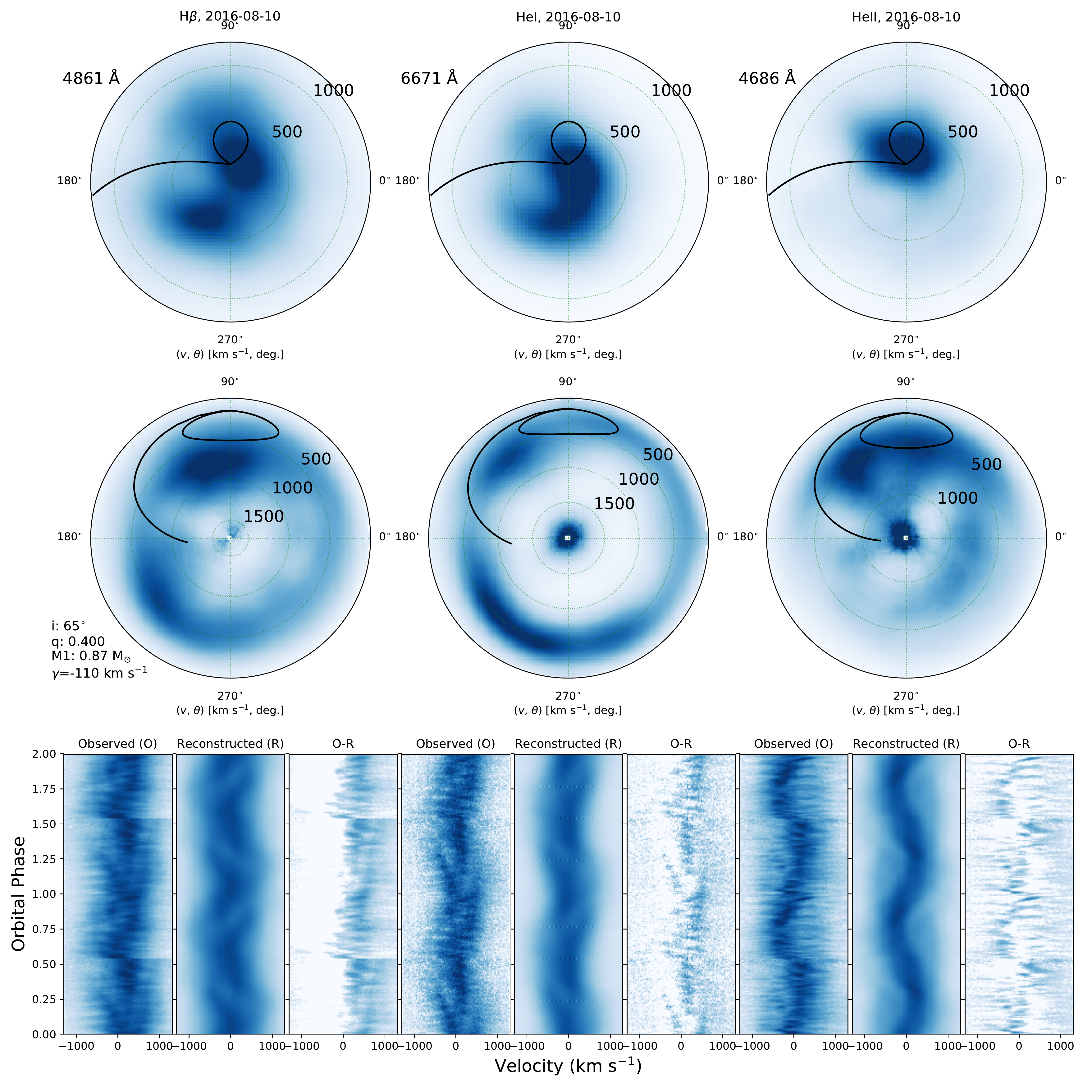}
    \caption{Doppler tomograms generated for H$\beta$ (left column), \ion{He}{I} (centre column), and \ion{He}{II} (right column). The top row shows the regular tomograms, the middle row shows the inverted tomograms, and the bottom row shows the observed spectra for the relevant line, the reconstructed spectra after generation of the Doppler tomogram, and the residual spectra after subtraction of the reconstructed spectra from the observed spectra. The position of the secondary star and ballistic stream are shown in black, and have been calculated using the binary parameters given in the lower left of each of the tomograms.}
    \label{fig:dopmaps}
\end{figure*}

Here, we took the \ion{He}{II} 4686\AA\ emission line, which has a single strong S-wave component, and constructed tomograms for a variety of systemic velocities, with the aim of minimising the residuals when the observed trailed spectra are subtracted from the model trailed spectra generated by the best fitting tomogram. We find that a systemic velocity of $\gamma=-110$ km s$^{-1}$ gives the best results with equal residuals in both negative and positive velocities, with an error of approximately 20 km s$^{-1}$. The residuals can be seen in lower right plot of Figure~\ref{fig:dopmaps}.

\subsubsection{Tomograms}
Doppler tomograms were then generated for H$\beta$, \ion{He}{I}, and \ion{He}{II} using a $\gamma=-110$ km s$^{-1}$. The results are shown in Figure~\ref{fig:dopmaps} with the binary system overlay calculated assuming an inclination of 65\degree, a mass ratio of 0.4, and a primary mass of 0.87 M$_{\odot}$.

Care must be taken when interpreting these tomograms as there are several periods shorter than the orbital period which affect the trailed spectra (mainly the beat and spin periods). As pointed out in \cite{Marsh1996}, the best way to analyse Doppler tomograms in the presence of shorter periods is to make Doppler tomograms at particular phases of one of the shorter periods. In their case, they created tomograms for several different values of the beat-phase, allowing them to create a ``stroboscopic'' view of the system. Such a detailed analysis on the spectra presented here is challenging, as the data only cover a single orbit, and the signal that dominates each line is not clear (as seen in Figure~\ref{fig:spectrograms} in which H$\alpha$ and \ion{He}{I} have power at both the spin and beat periods, depending on which part of the line is investigated).

Due to the location of the emission in the tomograms and the modulation of their data at the beat period of the system, \cite{Marsh1996} concluded that the observed emission was being caused by periodic illumination of both the inner face of the secondary star and the ballistic stream by the accreting poles of the WD. 

The tomograms presented here suggest a similar origin for the \ion{He}{II} emission during the low state - periodic illumination of the ballistic stream and the heated face of the secondary star, albeit at the half-beat period rather than the beat-period. This is best seen in the inverted tomogram, which shows the \ion{He}{II} emission originates from a region which is azimuthally extended around the secondary star and stream, and has a large radial extent.

The reconstructed spectra of both H$\beta$ and \ion{He}{I} contain valuable information regarding the line behaviour. In both of these lines, the reconstructed spectra show a broad emission component centred on 0 km s$^{-1}$ velocity alongside an S-wave absorption component which reaches maximum blue shift at phase 0.5 and maximum red shift at phase 1, which is completely lacking from the \ion{He}{II} spectra. This feature is also visible in the trailed spectra of H$\beta$ given in Figure 8 of \cite{Marsh1996} with exactly the same behaviour.

The inverted Doppler tomograms of these lines show evidence for the presence of a ring of material in the system, with an absorption dip occurring where the predicted ballistic stream crosses this ring (located at approximately at 165\degree). 

\subsubsection{Short Period Behaviour}\label{sec:2016_spin}

Along side the orbital modulation visible in the trailed spectra, there are also short pulses which extend to much higher velocities than the previously discussed S-wave components. This is perhaps best seen at $\phi\sim0.75$ in the \ion{He}{II} data shown in Figure~\ref{fig:trailed_heII}. Here the S-wave emission component has a width of $\sim$200 km s$^{-1}$, but the line itself extends from $-800$ km s$^{-1}$ to $+800$ km s$^{-1}$.

\begin{figure}
	\includegraphics[width=\columnwidth]{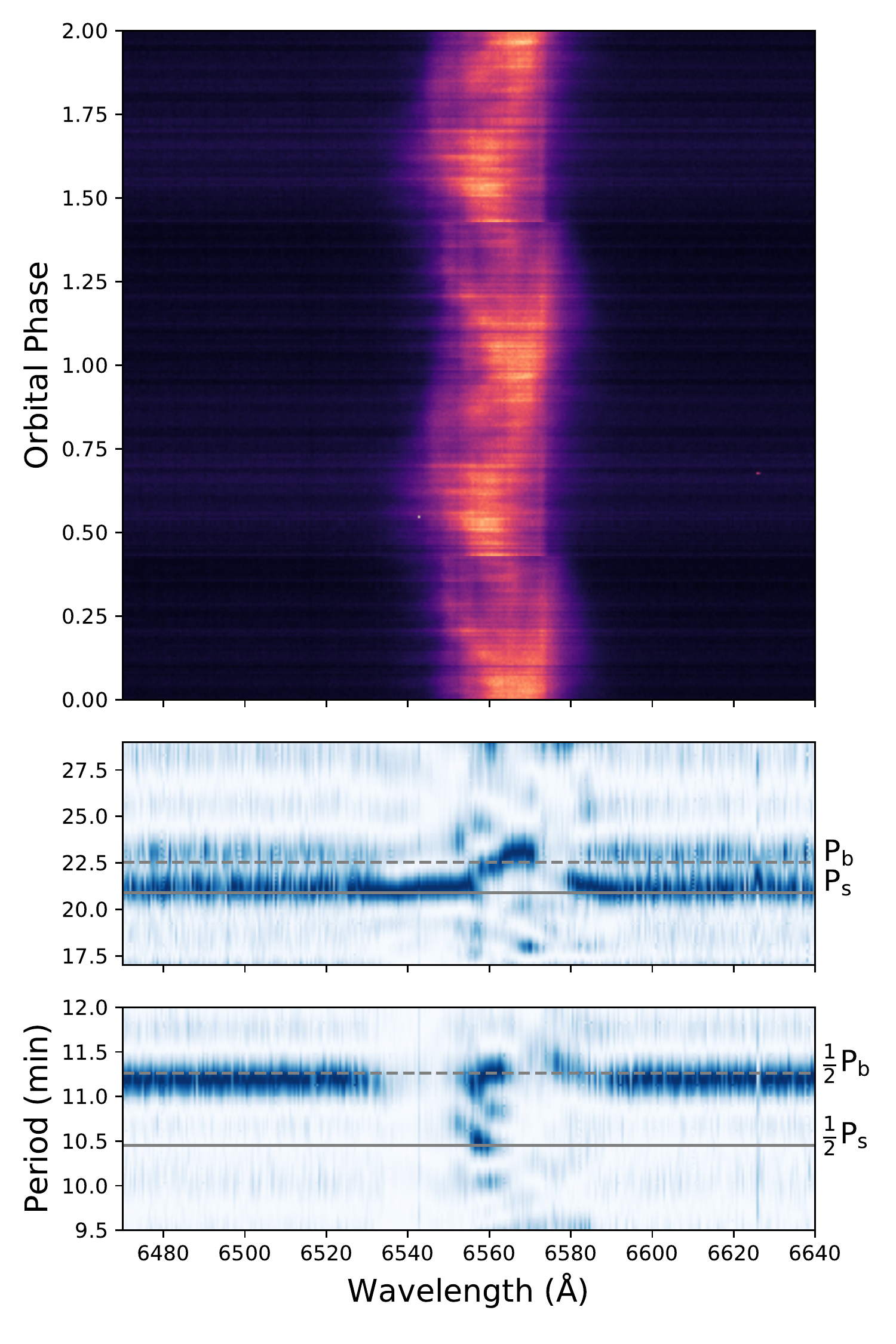}
    \caption{\textit{Top:} Trailed spectra around H$\alpha$ taken using the WHT during the 2016 low state. \textit{Middle:} A zoomed spectrogram focusing on the spin period of the WD ($P_S$) and beat period of the system marked ($P_B$). \textit{Bottom:} A zoomed spectrogram focusing on the half spin period of the WD (0.5$P_S$) and half beat period of the system marked (0.5$P_B$).}
    \label{fig:wht_spectrogram}
\end{figure}

A spectrogram of the continuum around H$\alpha$ during the low state is shown in Figure~\ref{fig:wht_spectrogram}. The continuum is clearly modulated at the spin period ($P_S$) of the WD and at the half beat period (0.5$P_B$) of the system, with residual power also at the beat period. This is consistent with optical photometry taken around these spectroscopic observations \citep{littlefield2016c}. However, in the blue wings of the H$\alpha$ emission feature, there is little power at half-beat and beat periods, while there is a very clear signal at the spin period of the WD. This is in stark contrast to the core of the emission line, where the strongest signal is at the beat and half-beat periods. This is visible in the first panel of Figure~\ref{fig:spectrograms}, which shows the spectrograms after continuum subtraction.

As noted in Section~\ref{sec:SpinBeat}, these spectrograms can be used to estimate which accretion mode was dominant during the low state. Perhaps the most important line and most easily interpreted for this purpose is \ion{He}{II} 4686 \AA. During the 2016 low state, the strongest period present in the spectrogram of this line was at the half beat period (11.29 min), with power also present at the beat period and no power at the spin or half-spin periods. The $V/R$ plot shown in Figure~\ref{fig:v_r} shows slightly different behaviour, with a very strong periodicity located at the spin period of the WD and lower amounts of power located between the half-spin and half-beat periods. This suggests that during these observations, there was a stationary component which showed varying flux levels with a period equal to the half-beat period, and another component which moved between negative and positive velocities with a period equal to that of the spin period.

The combination of the spectrogram and $V/R$ data suggests a particular origin for this emission feature. The stationary, flux varying component is likely to arise due to periodic reflection off or heating of a structure which is fixed in the binary rest frame - the bright spot at which the ballistic stream intersects the outer edge of the disc, the coupling point between this ballistic stream and the WDs magnetic field (in a stream-fed model) or, if the secondary is being significantly heated by the WD, then at the surface of the secondary star facing the WD (referred to hereafter as the inner face of the secondary). The component which led to the large $V/R$ signal at the spin period of the WD is likely to come from close to the WDs magnetic poles. The fact that the signal is at the spin period and not at the half spin period suggests that only the accretion structures linked to one of the magnetic poles is directly visible to the observer over the spin period.

Interpretation of the H$\alpha$ and \ion{He}{I} spectrograms is far more difficult due to the significant structure visible in both the trailed spectra and in the spectrograms. Both lines show pulsed emission at the spin period of the WD at blue wavelengths relative to the central wavelength, and both lines also have power at the beat period at red wavelengths relative to the central wavelength. The $V/R$ values for these lines are both modulated at the WD spin period, suggesting that the variability in this line is coming from the visible accretion structures attached to one of the WDs magnetic poles.

\subsubsection{Accretion geometry}
Analysis of the spectra taken during the 2016 low state challenges the conclusions that the dominant accretion mode during this period was stream fed as proposed by both \cite{littlefield2016c} and \cite{Kennedy2017}. The main conflict between the spectroscopy data and the photometry/X-ray data lies in the spectrogram and $V/R$ ratios for \ion{He}{II}. For the reasons laid out in Section~\ref{sec:2016_spin}, the spectroscopy suggests that the majority of \ion{He}{II} emission comes from illumination of either the coupling region between the disc and WDs magnetic field or from the heated inner face of the secondary by accretion onto both poles. Additionally, the spectroscopy suggests there is also a small amount of emission produced by material which is in an accretion disc and is coupling to the magnetic field, providing power at the WD spin period in the $V/R$ ratio.

 As noted above, the absorption feature visible in \ion{He}{I} was present with the same orbital phase during the high state, and has been attributed to absorption by material which has been ejected from the orbital plane around the impact region between the ballistic stream and the disc \citep{Hellier1990}. This is consistent with the tomograms shown in Figure~\ref{fig:trailed_heI}, and helps support the proposition that an accretion disc was still present around the WD at the time of these observations. Additionally, both H$\alpha$ and \ion{He}{I} lines show power at the spin period of the WD. The analysis of these data in combination with X-ray data taken during the low state which showed that accretion onto the WD was modulated at the half-beat period \citep{Kennedy2017} proves that FO Aqr was undergoing ``disc-overflow'' during this low state.

The occurrence of ``disc-overflow'' accretion at this time is unsurprising. At the lower accretion rates experienced during this time relative to the preceeding high state, the outer edge of the accretion disc (whose existence is hinted at in the above) will have shrunk. This means that the ballistic stream now penetrates further into the WDs Roche lobe before encountering disc, giving the material more time to expand vertically before encountering the disc. This, in combination with the higher velocity achieved by this infalling material and an increase in the radius at which the magnetic field begins to dominate over the ram pressure of the disc, will allow it to overflow the disc when it impacts the discs outer edge.

\subsection{2016 Recovery State} \label{sec:2016_recovery}
These spectra were taken after the system had recovered from the low state to the typical high state optical magnitude. The reason for referring to this state as the ``recovery'' state is that X-ray observations taken shortly after these optical spectra revealed that the system had still not fully recovered, suggesting there were still some difference between the accretion rate during these observations time and the typical high state accretion rate observed in previous works.

\subsubsection{Orbital Behaviour}
\begin{figure}
	\includegraphics[width=0.9\columnwidth]{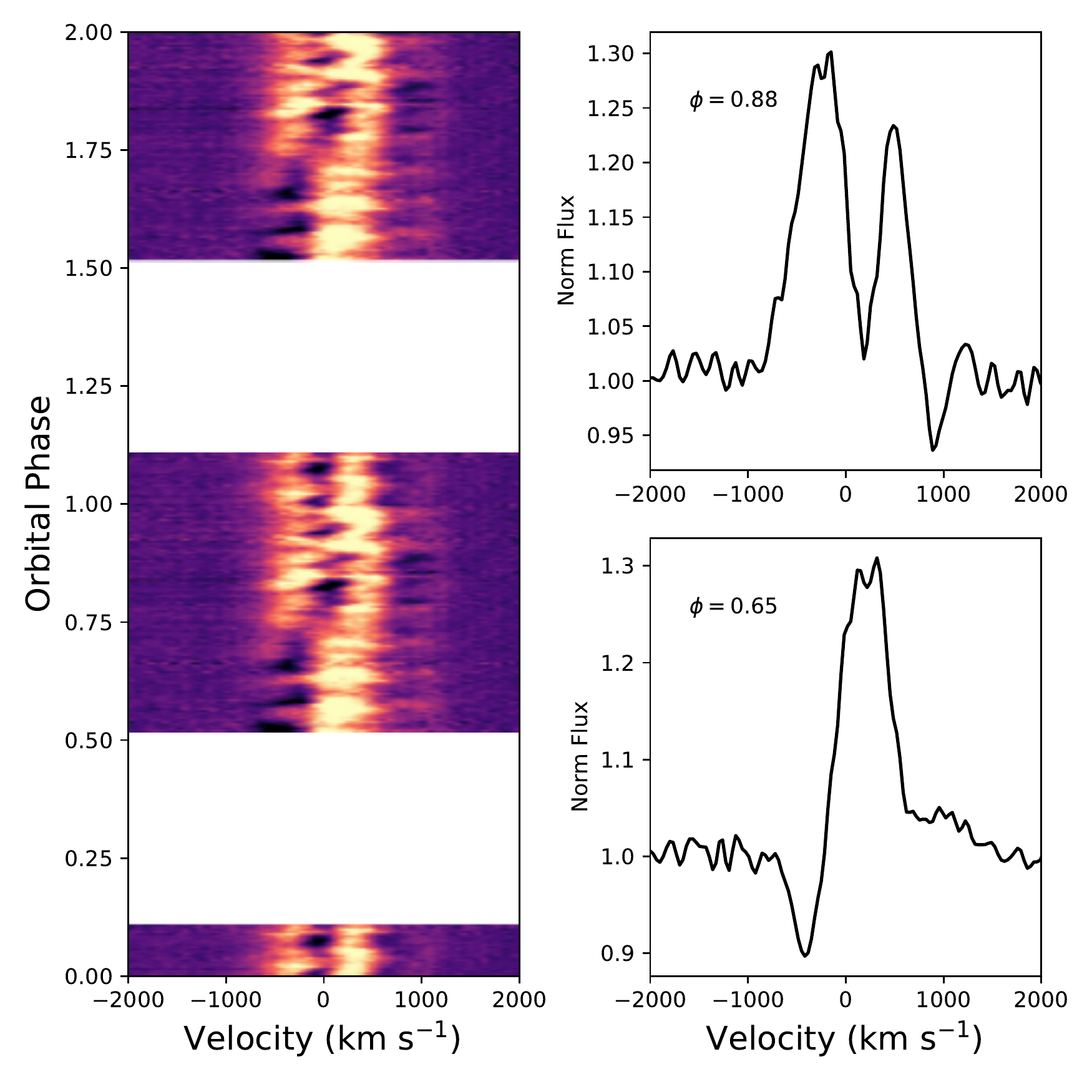}
    \caption{\textit{Left:} Trailed spectrum of \ion{He}{I} line plotted with a very high contrast to better highlight the absorption component which moves through the line. \textit{Right:} \ion{He}{I} profile at two different orbital phases, showing the movement of the absorption component.}
    \label{fig:LBT_HeI}
\end{figure}

The trailed spectra of H$\alpha$, \ion{He}{I}, and \ion{He}{II} during the recovery state are very similar to the spectra shown in \cite{Hellier1990} and \citet{Marsh1996}. This is particularly obvious if the high contrast trailed \ion{He}{I} plot in Figure~\ref{fig:LBT_HeI} is compared directly with the trailed spectra of H$\beta$ shown in Figure 8 of \citet{Marsh1996} (here we've chosen the \ion{He}{I} 4471 \AA\ line as it has a higher S/N than the 6678 \AA\ line). The minimum flux of the component which shows maximum blue shift at $\phi= 0.5$ is lower than the local continuum level, confirming that is not just the lack of an emission feature, but rather an absorption feature, which moves back towards the line centre between $\phi=0.5$ and $\phi=0.0$. Once the absorption component reaches the line centre ($\phi \sim 0.76$) an absorption feature also becomes detectable at high velocities. From this, we conclude that there is a strong emission component which spans from $-500$ km s$^{-1}$ to $+500$ km s$^{-1}$ that does not show much orbital variation, and an absorption component which moves with an S-wave like pattern through the emission feature over the orbit. The emission component also shows periodic flaring which extends to very high positive velocities ($>$ 1000 km s$^{-1}$).

The behaviour of \ion{He}{I} over the orbital period is very similar to that seen in \ion{He}{I} in the low state, where the full orbit of data shows that the strong central emission component moves very little over the orbital period, and an absorption S-wave moves through this line. This is perhaps best be seen by comparing the strong absorption feature that can be seen at centre of the line in both the middle-right plot and bottom-left plot of Figure~\ref{fig:trailed_heI}.

The H$\alpha$ line during the recovery state shows little motion over the half-orbit of data which is presented, and is dominated by pulsations. \ion{He}{II} shows an emission feature which follows an S-wave that is in phase with the S-wave observed in the low state and also in the data presented in \citet{Marsh1996}.

\subsubsection{Short Period Behaviour}

The spectrograms of the data taken during the recovery state (2nd column of Figure~\ref{fig:spectrograms}) are consistent with the emission lines varying at the spin and beat periods of the system, with very little power at shorter periods. This is supported by the Lomb-Scargle periodograms of the V/R values for each line (2nd column of Figure~\ref{fig:v_r}). The strongest periods in these periodograms are not located at exactly the spin or beat period, which is unsurprising given these observations only lasted for 3.1 hr.

The individual pulsations are easily visible in each line, and extend to very high velocities. In Figure~\ref{fig:LBT_HeI}, there are clear periodic pulsations in the line which extend above velocities of $+1000$ km s$^{-1}$, and are anti-correlated with the absorption feature which is seen out at these high velocities

\subsubsection{Accretion geometry}
The combination of a signal at the spin period of the WD in both the spectrogram of the \ion{He}{II} line and in the power spectrum of its $V/R$ ratio requires that there be a component in the line which is moving from positive to negative velocities at the spin period of the WD. This is likely an accretion curtain formed from coupling between the accretion disc and the WDs magnetic field. As such, we conclude that the dominant accretion mode during the recovery state was a disc-fed scenario. This is inline with X-ray observations taken around the time of these spectra \citep{Kennedy2017}.

\subsection{2017 low and 2018 failed low States} \label{sec:2017_low}

While not as deep as the 2016 low state, the spectra obtained during the two low states in 2017 and 2018 showed many of the same properties as those obtained in 2016.

\subsubsection{Orbital Behaviour}
\begin{figure}
    \begin{center}
	\includegraphics[width=0.75\columnwidth]{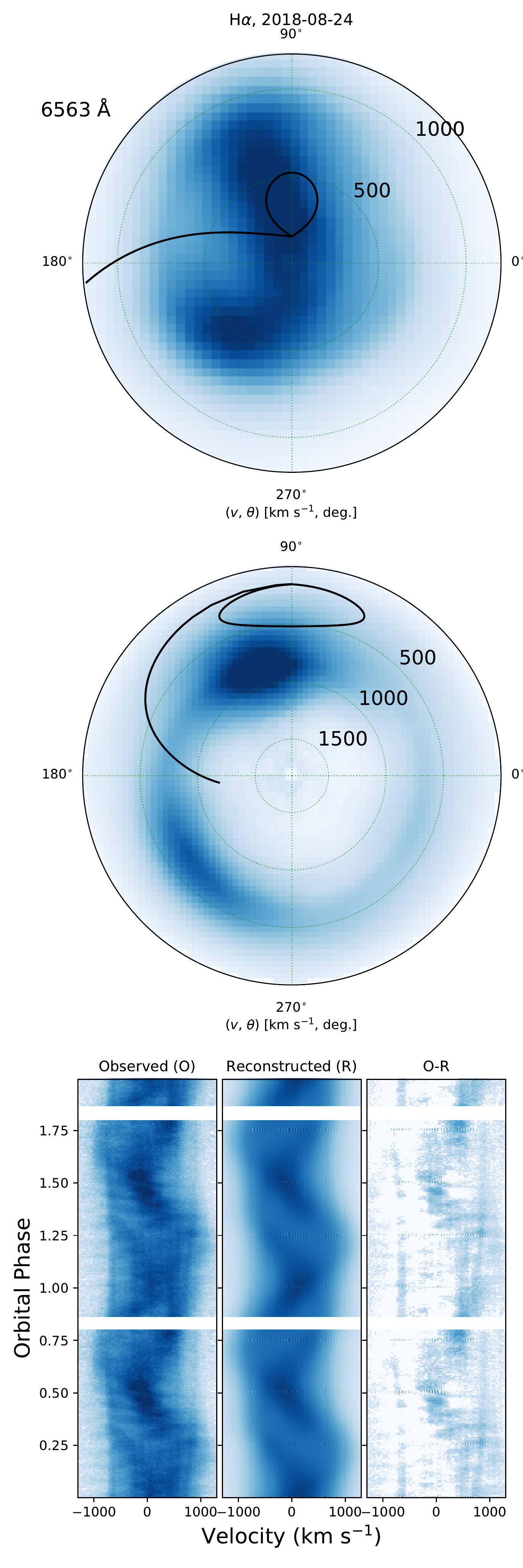}
	\end{center}
    \caption{Doppler tomograms generated for H$\alpha$. Individual plots are the same as those shown in Figure~\ref{fig:dopmaps}}
    \label{fig:NOT_dopmaps}
\end{figure}

The trailed spectra of H$\alpha$ and \ion{He}{I} taken during the 2017 and 2018 low states match the behaviour of those taken during the 2016 low state remarkably well, with the same components visible in both the H$\alpha$ and \ion{He}{I} lines. Most importantly, the absorption component remained present in \ion{He}{I}, and was in phase with all of the previous data, suggesting its location within the system had not changed. Figure~\ref{fig:NOT_dopmaps} shows a Dopper tomogram generated using the H$\alpha$ line during the 2018 low state. The features in the tomogram are identical to those seen in the tomograms of the Balmer and \ion{He}{I} lines in 2016.

\subsubsection{Short Period Behaviour}

Interpretation of the H$\alpha$ and \ion{He}{I} 2017 and 2018 spectrograms is difficult. In both years, both the H$\alpha$ and \ion{He}{I} show pulsed emission at the spin period of the WD at blue wavelengths relative to the central wavelength, similar to the behaviour seen during the 2016 low state. However no power is obvious at any shorter period.

During the 2017 low state, the $V/R$ values for both emission lines (2nd column from the right in Figure~\ref{fig:v_r}) are far more informative, showing very strong power at both the spin and beat periods. There is also evidence for power at the half-spin period in the $V/R$ data for the H$\alpha$ line. However, in 2018 (last column in Figure~\ref{fig:v_r}), the power spectra do not have any power at the beat period of the system, with the strongest peaks for both H$\alpha$ and \ion{He}{I} occurring at the spin period of the WD. 

\subsubsection{Accretion geometry}
The orbital behaviour of both the H$\alpha$ and \ion{He}{I} lines during 2017 low state bear more than a passing resemblance to the spectroscopy obtained during the 2016 low state. Combined with timing analysis of the H$\alpha$ and \ion{He}{I}, this suggests the rather complex ``disc-overflow'' accretion geometry was once again dominant during these observations.

During 2018, with no discernible power at the beat or half-beat periods in the $V/R$ plots for both lines, the accretion disc was likely present and feeding the WD . This is supported by photometry taken during this period, which shows the strongest periodicity in the light curve was that of the spin period \citep{littlefield19}. The lack of any signal at the beat period of the system is the main reason for assigning the ``failed'' low state label to this data, since it does not match the low state behaviour observed in 2016 and 2017, or the recovery state behaviour seen in 2016.

\section{Comparison of States} \label{sec:comparison}

\subsection{The cooler optical spectrum}
The average optical spectrum in the low and high states have much in common. Most apparent is that both spectra show the same ensemble of emission lines. The most substantial difference between the two states is the spectral index, which was measured by fitting a  power law ($F_{\rm \lambda} \propto \lambda^{\alpha}$) to the average spectrum in each state. Historically, the spectrum in the high state shows a sharp, distinct rise at shorter wavelengths with a spectral index of -2.3 (\citealt{Shafter1982}; \citealt{Martell1991}), and a UV spectrum measured with the \textit{Hubble Space Telescope}'s Faint Object Spectrograph matches this rise. This composite optical/UV spectrum has been fit with a composite model of two black bodies, one with a temperature of 36000 K coming from a small area that represents the WD, and another with a temperature of 12000 K coming from a larger area which dominates the optical spectrum and comes from a much larger area than the hot component, suggesting this cooler component is related to the accretion curtains (\citealt{1989ApJ...342..493C}; \citealt{Martino1999}).  In the recovery state, the spectral index was found to be $-(2.2\pm0.1)$, consistent with the value measured previously in the high state. The low state spectrum has a much less pronounced, but still distinguishable, rise towards shorter wavelengths, which is characterised by the flatter spectral index ($F_{\lambda} \propto \lambda ^{-(1.5\pm0.1)}$). 

In the high state, the 12,000 K black body component has been attributed to the temperature of the region where the accretion curtains meet the WD. If this is true, then the spectral index of the optical spectrum should vary over the spin period of the WD as this region passes into and out of our line of sight. While the power law of the recovery state ($F_{\lambda} \propto \lambda ^{-2.2}$) is consistent with the value measured in the high state, we cannot test if the spectral index varies over this period as the data were taken through changing levels of cloud cover, making the determination of the spectral index from any individual spectrum challenging.

A decrease in the temperature or in the size of this 12,000 K region would lower the flux from shorter wavelengths substantially more than from longer wavelengths, assuming the 36,000 K emission from the WD remains the same. This is the most likely cause in the change in spectral index between the high and low states. The decrease in either temperature or size of this region is most likely related to a reduced accretion rate, which would deplete the accretion disc and reduce the density of the coupling region. During future low states this may be investigated by obtaining UV spectroscopy of the source alongside optical data.

\subsection{Comparison of emission lines}\label{sec:comp_lines}
The shapes and behaviours of the various emission lines in the trailed spectra presented in this paper vary on a line-by-line basis between the low, failed low, and recovery states.

In the low state, the H$\alpha$ line has a complex, 3 component appearance, and extends out to high red velocities of approximately $+1000$ km s$^{-1}$ and $-1000$ km s$^{-1}$ respectively. In the trailed spectra, no variations from shorter periodicites are immediately obvious, but can be detected when the lines are analysed using either a Lomb Scargle periodogram or by measuring the change in flux at blue wavelengths over red wavelengths versus time (the $V/R$ ratio). In the recovery state, the line is less extended (between $\pm$500 km s$^{-1}$ and $\pm$700 km s$^{-1}$ depending on the orbital phase), and pulsations within the line are very obvious.

The change in the extent of the emission lines in radial velocity between the recovery state data and the low state data here lends strong support to the theory that the main accretion mode in FO Aqr was different in both states. \cite{Ferrario1999} predicted that for stream-fed systems, emission lines should have very large radial velocities due to the fact that we are seeing material follow field lines at essentially free-fall velocity, while the radial velocities of emission lines in a disc-fed system should be significantly lower (see their Figure 3 and 4). Comparing the recovery state data shown in Figure~\ref{fig:trailed_ha} with all of the other panels in the same Figure shows that the H$\alpha$ line in the recovery state had a much narrower radial velocity extent when compared to low state data.

The similarity of the lines (particularly H$\alpha$) between the 2016 low state, 2017 low state, and 2018 failed low state is surprising as FO Aqr did not fade to the same degree during 2017 (minimum V band magnitude $\sim$14.7) and 2018 (minimum V band magnitude $\sim$14.2) as it had in 2016 (minimum V band magnitude $\sim$15.7). Photometry of the 2017 and 2018 states is discussed in detail by \citet{littlefield19}.

Through out all of our observations, two components consistently appeared, and were in phase across each epoch. These two components are: i) the S-wave emission feature seen in \ion{He}{II} which has the same amplitude and phasing in both the low and high states; and ii) the S-wave absorption component in \ion{He}{I}. This means that either these two regions are unaffected by any change in the mass accretion rate, or any effects that occur during the deepest part of the low states are short lived, with the lines quickly returning to their standard behaviour.

\begin{figure}
	\includegraphics[width=1.0\columnwidth]{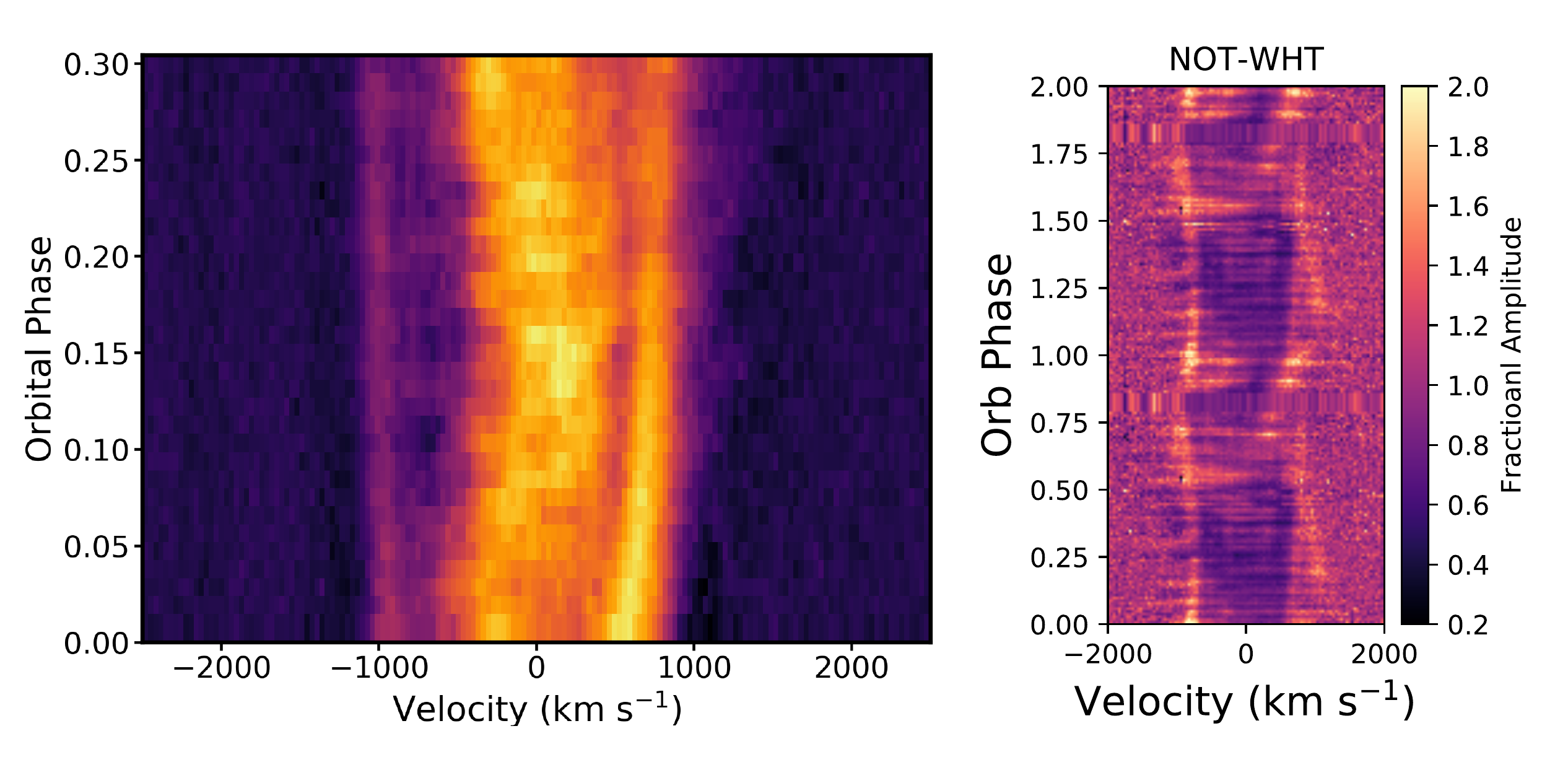}
    \caption{\textit{Left:} Time-resolved H$\alpha$ emission from MDM taken 2016 July 10. An unsharp mask has been applied to the image to reduce the brightness of the central emission. Note the slowly varying features at $-1000$~\kms\ and $+800$~\kms.  \textit{Right:} The difference between the 2018 July NOT stacked spectrum and the 2016 August WHT spectral sequence. The residuals show slowly varying high-velocity features during the 2018 low state.  }
    \label{fig:high_vel}
\end{figure}

\subsection{Evidence for outflow in the low state}

Bipolar outflows and jets are ubiquitous in accreting stellar systems ranging from young stellar objects to neutron star and black hole X-ray binaries \citep{livio2000}. While high accretion rate CVs are well known to have weakly collimated disc winds (see \citealt{2015MNRAS.450.3331M} and references within), highly collimated bipolar outflows and jets in accreting WD systems have been harder to identify at optical wavelengths \footnote{Note that radio emission from dwarf novae CVs during outbursts may be produced by a transient jet (\citealt{2008Sci...320.1318K}; \citealt{2016MNRAS.463.2229C}; \citealt{2017MNRAS.467L..31M}).}. Satellite emission features attributed to collimated jets have been seen in a handful ``supersoft'' X-ray sources (SSSs) \citep{crampton1996,southwell1996,Becker98}. SSSs are thought to result from a high accretion rate on to a WD leading to continuous or episodic hydrogen burning at the surface. Searches for jets from classical CVs have failed to find clear evidence of satellite emission features \citep{hillwig2004}. Additionally, \citet{kh04} have proposed that outflows in CVs might cause the He I $\lambda 5876, 7063$ triplets to preferentially show P~Cyg compared to other He I lines.
One of the most striking features of the FO~Aqr low-state spectra are the blue and red shifted H$\alpha$ emission components with velocities between 600~\kms\ and 1000~\kms (the left and right hand side plots of Figure~\ref{fig:high_vel}). The features are also visible in the O-R panel of Figure~\ref{fig:NOT_dopmaps}. These high-velocity satellite features are not visible in the bright state and and are not always clearly detected in the low-state spectra. No P~Cyg profiles are seen for He I $\lambda 5876, 7063$ \citep{kh04}, although our spectra do not always cover those two features.

\begin{figure}
	\includegraphics[width=0.9\columnwidth]{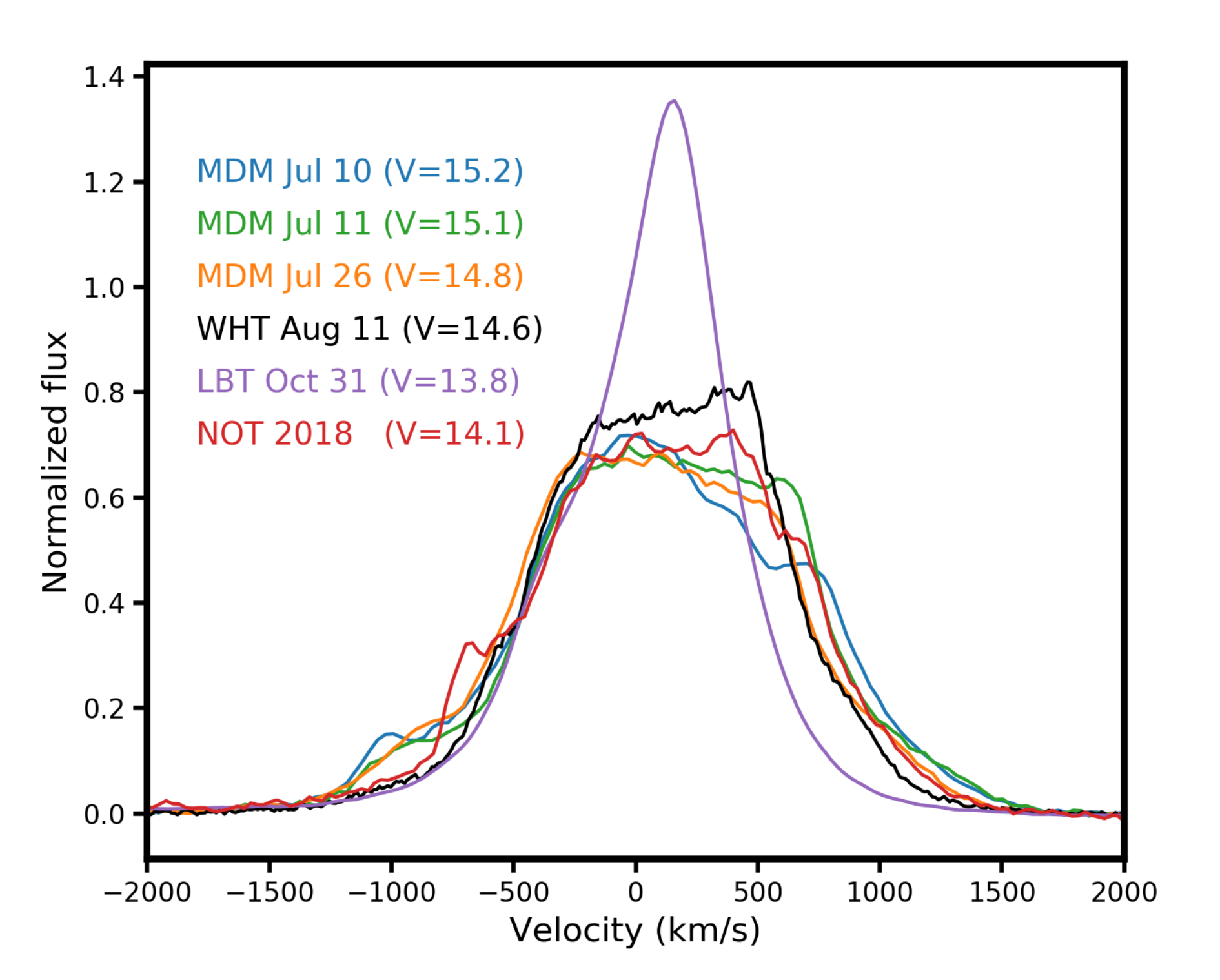}
    \caption{The average H$\alpha$ profile between orbital phases 0.2 and 0.3 for five epochs in 2016 and one in 2018. The highest velocities of the emission wings are generally seen when the system is the faintest. The line profile from the LBT spectrum obtained during the recovery state is substantially different from low-state profiles. }
    \label{fig:halpha}
\end{figure}

The satellite emission features shown in Figure~\ref{fig:high_vel} are similar to those seen in SSSs, which have been attributed to collimated jets. Satellite emission features in both FO~Aqr and SSSs are transient and episodic suggesting that the degree of the jet collimation can vary over days or weeks. As seen in Figure~\ref{fig:halpha}, the velocity of the satellite features (the small bumps at high velocities in spectra taken during a low state) in FO~Aqr tends to roughly inversely correlate with the system luminosity.

We estimate the equivalent width (EW) of the blue-shifted satellite emission feature seen in 2016 July to be $-6$~\AA. The EW of the red-shifted feature is larger by a factor of two, although its exact value is strongly dependent on how the main H$\alpha$ emission is accounted for. The strengths of the satellite features in FO~Aqr are larger than those observed in SSSs which, when detected, have EW between $-1$~\AA\ and $-3$~\AA\ \citep{southwell1996,Becker98}.

For an orbital inclination of 65$^\circ$, as implied by the disc eclipse, the full outflow velocity is more than double the velocity implied by the observed emission line shifts. When present, the satellite emission features show a slow velocity variation of about 100~\kms over the orbital period, or about 10\%\ of the velocity amplitude. This implies that the outflow direction that is nearly perpendicular to the orbital plane.

\citet{knigge1998} derive a scaling relation for the detectability of satellite features produced by jets in accreting WDs. They suggest that the equivalent width of jet emission features should scale with the mass accretion rate on to the WD. We should, therefore, be less likely to see jet features in low states of disc dominated CVs as fainter discs suggest decreased mass transfer rates. This is in direct conflict with our results. However, there are several important caveats to consider. The results of \citet{knigge1998} assume the WD is accreting exclusively from a disc, with the magnetic field of the WD assumed to lead to negligible accretion at the magnetic poles. The IP classification of FO~Aqr implies that its WD has a fairly strong magnetic field, meaning a direct comparion with previous results is difficult. Considering jets are mainly driven by a hydromagnetic process \citep{livio2000}, then it may not be surprising that FO~Aqr is more likely to drive collimated outflows than ordinary CVs.

In addition, \citet{littlefield19} have shown that the WD in FO~Aqr has recently switched from spinning-up to spinning-down and that low-states appear correlated with times when the WD spin rate is slowing. The WD spin-down is a source of power equivalent to about a Solar luminosity, yet the optical light from the system has declined from its historical average. The spin-down of the WD may be the additional energy source needed to drive collimated outflows as suggested by \citet{livio2000}. 

As previously mentioned, the accretion discs in high accretion rate CVs are known to produce winds. While there is evidence of a disc being present during the low states observed in FO Aqr, we find this is unlikely to be the case here for the following reason. During the normal high state, the accretion disc is much larger and hotter than during the low state. As such, the system should be more capable of producing a disc wind in the high state than in the low state. The lack of evidence for a disc wind during the system's usual high state suggests that a disc wind is not responsible for the apparent outflow proposed during the low state.

The existence of the proposed outflow can be tested with future optical and ultraviolet spectroscopy and radio observations of FO~Aqr during future low states.

\subsection{Implications for the fate of the disc}

Since observations of the first low state in FO Aqr, there has been detailed discussion as to whether an accretion disc is permanently present in the system, or whether it disappeared during the deepest part of the low state and was rebuilt as the system recovered (\citealt{littlefield2016c}; \citealt{2017A&A...606A...7H}; \citealt{littlefield19}). The current consensus is that the accretion disc should have dissipated during the low state. 

The evolution of the central component of the H$\alpha$ line is consistent with this scenario. During the low state, the motion of the line centre is modulated predominantly at the orbital period and shows maximum blueshift near $\phi_{orb} = 0.6$, similar to the behaviour of the ballistic accretion stream in polars. Although the velocity amplitude is lower than in polars, this might be caused by the presence of multiple line components \citep[e.g., as in the discless IP V2400 Oph;][]{Hellier2002}, as well as dilution of the line wings by the relatively strong continuum. \citet{2017A&A...606A...7H} also noted that during the low state, there may still be material which is impacting the magnetosphere and sustaining a structure within the orbital plane which resembles a traditional accretion disc. There is evidence for this structure in both the spectrograms and the $V/R$ plots of H$\alpha$ and \ion{He}{I}, which show strong modulation at the spin period of the WD during the low states. Since such a signal could only be coming from coupling between the WDs magnetic field and an accretion disc, we find it likely that a residual disc still existed in the system during the observations. This coupling region also contributes to the \ion{He}{II} emission.

Based on power spectral analysis of optical photometry, \citet{littlefield2016c} concluded that there was a direct interaction between at least part of the accretion flow and the WDs magnetosphere during the low state. Specifically, they invoked the stream-overflow model, in which part of the ballistic stream survives its initial collision with the disc and remains on a ballistic trajectory until it strikes the magnetosphere. The phase dependant absorption in H$\alpha$ and \ion{He}{I} support this theory as it is in phase with the expected absorption from overflowing gas at the impact region screening parts of the accretion disc.

\section{Conclusions}\label{sec:conc}

The spectra presented here paint a complicated picture for the accretion structures within FO Aqr. We find evidence for the existence of an accretion disc in each set of observations, and the absorption S-wave which has been seen in previous studies persists across all our data, reaching maximum blue and red shifts at the same orbital phase regardless of the systems brightness. There are also many differences between the various observations, suggesting that the dominant accretion mode varies on time scales of $\sim$ months.

In particular, we find that the following accretion regimes were dominant between 2016-2018.

\begin{description}
\item \textbf{2016 Low State:} The system was likely undergoing ``disc-overflow'' accretion, and the accretion structures at both poles were contributing to the periodicties seen in photometry and spectroscopy around this time. This is best highlighted by the behaviour of the \ion{He}{II} line.
\item \textbf{2016 Recovery State:} The system was undergoing ``disc-fed'' accretion, with little evidence for accretion due to ``disc-overflow''.
\item \textbf{2017 Low State:} The system returned to a ``disc-overflow'' mode, with the disc again depleted enough such that accretion structures feeding both poles were visible.
\item \textbf{2018 Failed Low State:} The system was undergoing ``disc-fed'' accretion. The disc had dissipated slightly but was not as depleted as it was during the 2016 and 2017 low states.
\end{description}

There are still many open questions in relation to FO Aqr. Future high resolution spectroscopy during another low state is highly encouraged to better explore the evolution of the disc as it dissipates and reforms, alongside looking for more convincing evidence of a transient outflow in the system. More general currently unanswered questions also exist, such as what is the mass of the WD, and what is the spectral type of the companion? The answers to all of these questions can be investigated by continued optical monitoring of FO Aqr, and obtaining spectroscopy at opportune moments.

\section*{Acknowledgements}

We would like to thank Daniel Mata Sanchez, Colin Clark, and Guillaume Voisin for providing comments on an early version of this paper. MRK acknowledges support from the Royal Society in the form of a Newton International Fellowship (grant number NF171019). MRK, PC, and PMG also acknowledge financial support from the Naughton Foundation, Science Foundation Ireland, and the UCC Strategic Research Fund which funded MRK during the start of this project. MRK and PMG acknowledge support for program number 13427 which was provided by NASA through a grant from the Space Telescope Science Institute, which is operated by the Association of Universities for Research in Astronomy, Inc., under NASA contract NAS5-26555. R.P.B. acknowledges support from the European Research Council under the European Union's Horizon 2020 research and innovation programme (grant agreement no. 715051; Spiders).

MRK would also like to thank the many friends and family who provided support throughout the course of preparing this paper.

The data presented here were obtained in part with ALFOSC, which is provided by the Instituto de Astrofisica de Andalucia (IAA) under a joint agreement with the University of Copenhagen and NOTSA. This paper used data obtained with the MODS spectrographs built with funding from NSF grant AST-9987045 and the NSF Telescope System Instrumentation Program (TSIP), with additional funds from the Ohio Board ofRegents and the Ohio State University Office ofResearch. The LBT is an international collaboration among institutions in the United States, Italy and Germany. The LBT Corporation partners are: The University of Arizona on behalf of the Arizona university system; Istituto Nazionale di Astrofisica, Italy; LBT Beteiligungsgesellschaft, Germany, representing the Max Planck Society, the Astrophysical Institute Potsdam, and Heidelberg University; The Ohio State University; The Research Corporation, on behalf of The University of Notre Dame, University of Minnesota and University ofVirginia. This work is based on observations obtained at the MDM Observatory, operated by Dartmouth College, Columbia University, Ohio State University, Ohio University, and the University of Michigan. The William Herschel Telescope is operated on the island of La Palma by the Isaac Newton Group of Telescopes in the Spanish Observatorio del Roque de los Muchachos of the Instituto de Astrofísica de Canarias. 

STSDAS and PyRAF are products of the Space Telescope Science Institute, which is operated by AURA for NASA. This research made use of Astropy, a community-developed core Python package for Astronomy \citep{2013A&A...558A..33A}.




\bibliographystyle{mnras}
\bibliography{FOAQr2} 




\appendix

\section{Flux and Equivalent Width measurements of spectral lines}

We measured the flux and equivalent width (EW) of various emission lines from the average spectrum of each night. These values are given in Table~\ref{tab:ew_flux_tab}. While the flux of many lines varied substantially between all observations taken during July 2016, the EW were all in near agreement during this low state, and the flux variations were due to poor and variable weather conditions in July 2016. This table also shows that the equivalent widths of the lines reached their highest values during the low states. Additionally, the EW values shown in Table~\ref{tab:ew_flux_tab} reveal that the Balmer decrement, H$\alpha$:H$\beta$, was 1.4:1 during the low state. This is lower than during the regular high state when the Balmer decrement was closer to 1.8:1, suggesting a significant change in the ionisation fraction in the low state.

\begin{table*}
	\centering
	\caption{Flux and EW for various emission lines in the optical spectrum. The errors are significantly higher for the LBT data due to clouds that severely affected many of the observations, and the values presented here are the mean of the values over the run, while the errors are the standard deviation of the values. There are no EW values for \ion{He}{II} from 2017-09-17 and 2018-07-24 as the spectra only covered H$\alpha$ and \ion{He}{I}. Missing flux values are marked with a -, whether it's due to limited wavelength range or the lack of a flux standard to calibrate the data.}
	\begin{tabular}{r c c c c c c c c}
		\hline
                                            &                               & Flux                              &                                   &                                       &               & E.W.          &               &             \\
        Date                                & H$\alpha$                     & H$\beta$                          & \ion{He}{I}                       & \ion{He}{II}                          & H$\alpha$     & H$\beta$      & \ion{He}{I}   & \ion{He}{II}\\    
		 				                    & erg cm$^{-2}$ s$^{-1}$        & erg cm$^{-2}$ s$^{-1}$            & erg cm$^{-2}$ s$^{-1}$            & erg cm$^{-2}$ s$^{-1}$                & \AA           & \AA           & \AA           & \AA\\
		\hline\hline
		2016-07-10                          & (5.9$\pm$0.5)$\times10^{-13}$ & (4.0$\pm$0.4)$\times10^{-13}$     & (7.7$\pm$0.6)$\times10^{-14}$     & (1.7$\pm$0.3)$\times10^{-13}$         & 230$\pm$30    & 110$\pm$10    & 34$\pm$5      & 39$\pm$4 \\ 
		2016-07-11                          & (2.3$\pm$0.5)$\times10^{-13}$ & (1.5$\pm$0.5)$\times10^{-13}$     & (3.1$\pm$0.7)$\times10^{-14}$     & (0.7$\pm$0.2)$\times10^{-13}$         & 240$\pm$30    & 100$\pm$20    & 38$\pm$6      & 39$\pm$6 \\
		2016-07-16                          & (7.1$\pm$0.5)$\times10^{-13}$ & (5.2$\pm$0.4)$\times10^{-13}$     & -                                 & (2.3$\pm$0.4)$\times10^{-13}$         & 220$\pm$20    & 120$\pm$10    & -             & 40$\pm$6 \\
		2016-07-26                          & (3.4$\pm$0.6)$\times10^{-13}$ & (2.4$\pm$0.6)$\times10^{-13}$     & (4.6$\pm$1.0)$\times10^{-14}$     & (1.2$\pm$0.3)$\times10^{-13}$         & 200$\pm$20    & 90$\pm$10     & 28$\pm$4      & 40$\pm$5 \\
		2016-07-29                          & (1.9$\pm$0.7)$\times10^{-13}$ & (1.1$\pm$0.6)$\times10^{-13}$     & (2.1$\pm$0.9)$\times10^{-14}$     & (0.5$\pm$0.2)$\times10^{-13}$         & 220$\pm$20    & 110$\pm$20    & 29$\pm$6      & 39$\pm$8\\
		2016-08-10                          & -                             & -                                 & -                                 & -                                     & 150$\pm$30    & 90$\pm$20    & 24$\pm$5      & 36$\pm$6\\
		2016-10-31                          & (11$\pm$6)$\times10^{-13}$    & (6$\pm$3)$\times10^{-13}$         & (8$\pm$5)$\times10^{-14}$         & (4$\pm$2)$\times10^{-13}$             & 80$\pm$14     & 25$\pm$4      & 6$\pm$1       & 14$\pm$2\\
		2017-09-17                          & -                             & -                                 & -                                 & -                                     & 170$\pm$20    & -             & 20$\pm$3      & -\\
		2018-07-24                          & -                             & -                                 & -                                 & -                                     & 150$\pm$30    & -             & 15$\pm$4      & -\\
		\hline
	\end{tabular}
	\label{tab:ew_flux_tab}
\end{table*}


\bsp	
\label{lastpage}
\end{document}